\documentclass[prl,twocolumn,floatfix,superscriptaddress,amsmath,aps,longbibliography]{revtex4-1}
\pdfoutput=1
\usepackage{dcolumn,graphicx,color,booktabs,microtype,afterpage}
\usepackage[colorlinks,plainpages=false,linkcolor=blue,urlcolor=blue,citecolor=blue,pdfpagemode=UseNone,pdfstartview=FitBH]{hyperref}
\usepackage{graphicx}
\usepackage{bmpsize}
\usepackage{amsfonts}
\usepackage{gensymb}
\usepackage{bm}

\begin{document}
\title{Field-Induced Spin Excitations in the Spin-1/2 Triangular-Lattice Antiferromagnet CsYbSe$_2$}
\author{Tao Xie}
\thanks{These authors contributed equally to this work}
\affiliation{Neutron Scattering Division, Oak Ridge National Laboratory, Oak Ridge, TN 37831, USA}
\author{Jie Xing}
\thanks{These authors contributed equally to this work}
\affiliation{Materials Science and Technology Division, Oak Ridge National Laboratory, Oak Ridge, Tennessee 37831, USA}
\author{S.~E.~Nikitin}
\affiliation{Paul Scherrer Institut, CH-5232 Villigen, Switzerland}
\author{S.~Nishimoto}
\affiliation{Department of Physics, Technical University Dresden, 01069 Dresden, Germany}
\affiliation{Institute for Theoretical Solid State Physics, IFW Dresden, 01069 Dresden, Germany}
\author{M.~Brando}
\affiliation{Max Planck Institute for Chemical Physics of Solids, N\"{o}thnitzer Str. 40, D-01187 Dresden, Germany}
\author{P.~Khanenko}
\affiliation{Max Planck Institute for Chemical Physics of Solids, N\"{o}thnitzer Str. 40, D-01187 Dresden, Germany}
\author{J.~Sichelschmidt}
\affiliation{Max Planck Institute for Chemical Physics of Solids, N\"{o}thnitzer Str. 40, D-01187 Dresden, Germany}
\author{L.~D.~Sanjeewa}
\affiliation{Materials Science and Technology Division, Oak Ridge National Laboratory, Oak Ridge, Tennessee 37831, USA}
\author{Athena~S.~Sefat}
\affiliation{Materials Science and Technology Division, Oak Ridge National Laboratory, Oak Ridge, Tennessee 37831, USA}
\author{A.~Podlesnyak}
\affiliation{Neutron Scattering Division, Oak Ridge National Laboratory, Oak Ridge, TN 37831, USA}

\begin{abstract}
A layered triangular lattice with spin-1/2 ions is an ideal platform to explore highly entangled exotic states like quantum spin liquid (QSL). Here, we report a systematic in-field neutron scattering study on a perfect two-dimensional triangular-lattice antiferromagnet, CsYbSe$_2$, a member of the large QSL candidate family rare-earth chalcogenides. The elastic neutron scattering measured down to 70 mK shows that there is a short-range 120$\degree$ magnetic order at zero field. In the field-induced ordered states, the spin-spin correlation lengths along the $c$ axis are relatively short, although the heat capacity results indicate long-range magnetic orders at 3 T $-$ 5 T. The inelastic neutron scattering spectra evolve from highly damped continuum-like excitations at zero field to relatively sharp spin wave modes at the plateau phase. Our extensive large-cluster density-matrix renormalization group calculations with a Heisenberg triangular-lattice nearest-neighbor antiferromagnetic model reproduce the essential features of the experimental spectra, including continuum-like excitations at zero field, series of sharp magnons at the plateau phase as well as two-magnon excitations at high energy. This work presents comprehensive experimental and theoretical overview of the unconventional field-induced spin dynamics in triangular-lattice Heisenberg antiferromagnet and thus provides valuable insight into quantum many-body phenomena.
\end{abstract}
\maketitle

Geometrically frustrated magnets provide an intriguing playground for investigation of the novel phenomena in condensed matter physics~\cite{Diep,balents2010spin,rau2019frustrated}.
Strong frustrations may produce large degeneracy of the ground state and prevent formation of magnetic order in favor of exotic states such as quantum spin-liquid (QSL) or spin-ice phases~\cite{balents2010spin,rau2019frustrated,Broholmeaay0668}.

Two-dimensional (2D) triangular-lattice antiferromagnet (TLAF) with nearest-neighbor (NN) isotropic antiferromagnetic (AF) coupling is a prototypical example of frustrated antiferromagnet. In classical regime ($S\rightarrow\infty $) TLAF orders into a long-range 120$\degree$ AF state at 0~K~\cite{Starykh2015}, while the ground state of TLAF in the quantum limit ($S$~=~1/2) was proposed to be a QSL~\cite{Anderson, fazekas1974ground}. Further theoretical and numerical investigations have shown that $S$~=~1/2 TLAF also shows the 120$\degree$ AF ground state, although the ordered moment is reduced as compared to the classical case, due to strong quantum fluctuations~\cite{bernu1992signature, white2007neel}.
The quantum fluctuations also favor collinear up-up-down (UUD) order and stabilize a magnetization plateau at $M_s/3$ ($M_s$ is saturation magnetization) for $S$~=~1/2 TLAF in a finite magnetic field regime~\cite{kamiya2018nature,Xing20192,Ranjith2019naybse,ranjith2019field,chubokov1991quantum, yamamoto2014quantum,SCHMIDT20171}.
Generalization of this model to the case of anisotropic exchange interactions drastically enriches the magnetic field-temperature ($B$-$T$) phase diagram and produces a number of novel quantum phases including the QSL state for certain range of parameters~\cite{Liyaodong2016,zhu2018topography,maksimov2019anisotropic}, although the experimental realizations of clean TLAF remain rare.

\begin{figure}[tb]
\center{\includegraphics[width=1\linewidth]{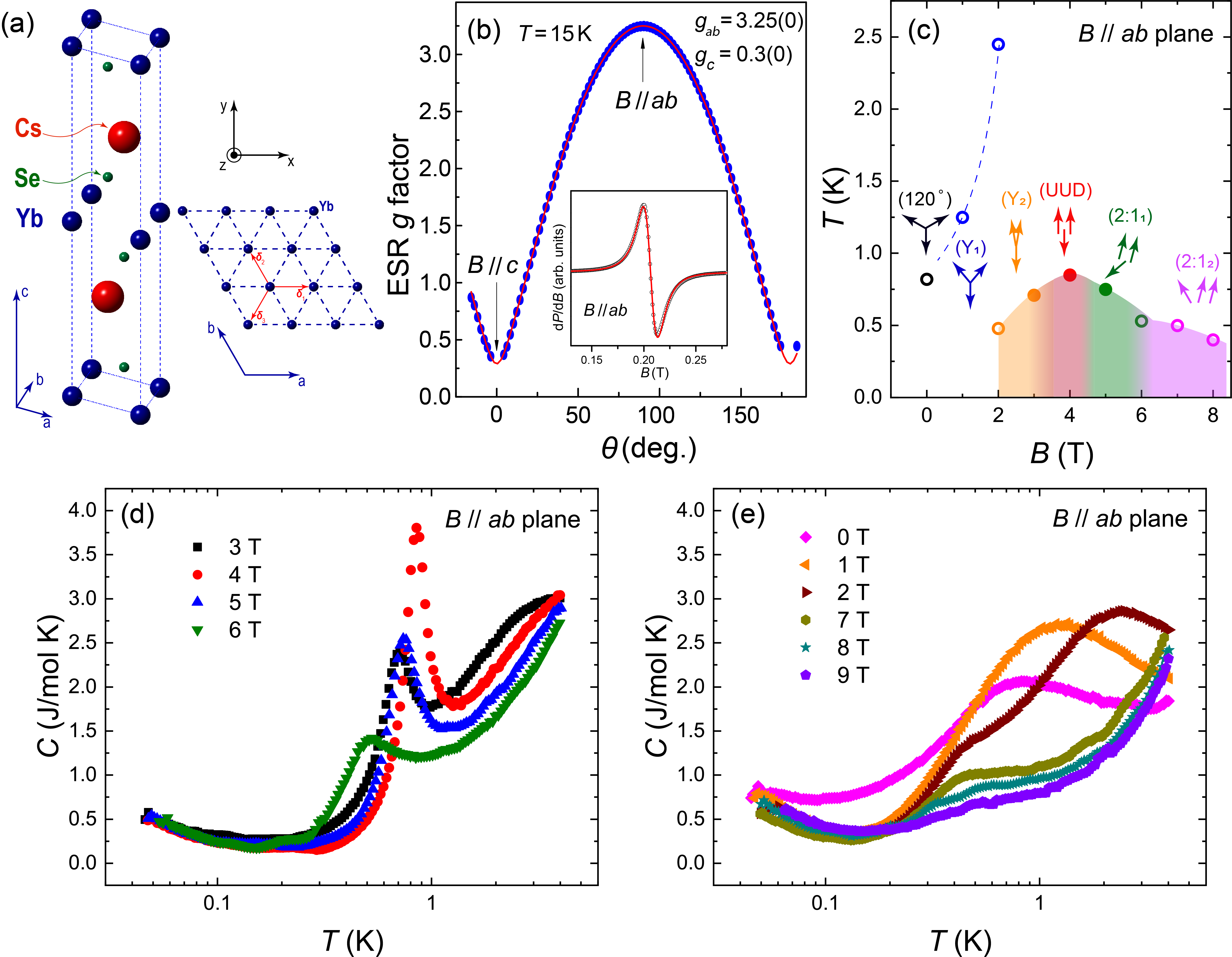}}
  \caption{(a) Crystal structure and the Yb triangular layer of CsYbSe$_2$. (b) Anisotropy of the ESR $g$ factor at 15 K, the solid line indicates $g(\theta)$ = $\sqrt{g_{ab}^2\sin^2{\theta}+g_{c}^2\cos^2{\theta}}$ with $g_{ab}$ = 3.25(0) and $g_c$ = 0.3(0). The inset is a typical ESR spectra measured with $B$ $\parallel$ $ab$, the red solid line in the inset is a Lorentzian line fitting. (c) The schematic phase diagram extracted from the heat capacity results. (d)--(e) Temperature dependence of the heat capacity measured at different magnetic fields with $B$ $\parallel$ $ab$.
  }
  \label{fig1}
\end{figure}

Recently, Yb-based delafossites $A$Yb$Q_2$ ($A$ = Li, Na, K, Rb, Cs, Tl, Ag, Cu; $Q$ = O, S, Se, Te) [Fig.~\ref{fig1}(a)] attracted widespread attention in search of QSL candidates~\cite{liu2018rare,ranjith2019field, Ranjith2019naybse,Xing20191,ding2019gapless, ma2020spin, ferreira2020frustrated, liybs2, iizuka2020single, bordelon2019field, baenitz2018naybs,sichelschmidt2020effective,dai2020spinon}.
Combination of strong spin-orbit coupling and crystalline electric field creates an effective spin $S$~=~1/2 at each Yb site at low temperatures due to Kramers degeneracy of a ground state doublet~\cite{baenitz2018naybs,sichelschmidt2020effective,Zhang2021CEF,dai2020spinon}.
No long-range order was observed at zero field in $A$Yb$Q_2$ by different measurement techniques, including heat capacity, magnetization, muon spectroscopy and neutron diffraction~\cite{liu2018rare, ranjith2019field, Ranjith2019naybse, Xing20192, ding2019gapless, ma2020spin, ferreira2020frustrated, liybs2, iizuka2020single, bordelon2019field}.
The Yb ions form perfect triangular layers [Fig.~\ref{fig1}(a)] without any intrinsic structural disorder, in contrast to another well-known TLAF QSL candidate, YbMgGaO$_4$, which shows valuable Mg/Ga site disorders~\cite{li2015gapless, li2015rare, li2016muon, shen2016evidence, paddison2017continuous, zhang2018hierarchy, li2017crystalline, steinhardt2020constraining, li2019ybmggao4, shen2018fractionalized, zhu2017disorder, zhu2018topography, kimchi2018valence}.
Phase diagrams of NaYbSe$_2$ and NaYbO$_2$~\cite{Ranjith2019naybse,bordelon2019field} include a QSL state and magnetic field-induced ordered states, that makes $A$Yb$Q_2$ an ideal platform to study these different quantum phases in $S$~=~1/2 TLAF.

Low-energy spin dynamics of the Yb-based delafossites have been studied by single-crystal inelastic neutron scattering (INS) at zero field in CsYbSe$_2$ and NaYbSe$_2$~\cite{Xing20192,dai2020spinon}.
The signatures of the gapless excitation continuum, whose intensity concentrates around K point, $i.e.$, (1/3, 1/3, 0) of Brillouin zone (BZ) and extends up to 2.5~meV, were found in both materials suggesting a QSL ground state with a spinon Fermi surface~\cite{dai2020spinon}.
However, very little is known about spin dynamics of the field-induced phases in these materials, because all previous in-field INS measurements were limited by polycrystalline samples~\cite{bordelon2019field, bordelon2020spin}.
Recent INS study of spin excitations in Ba$_3$CoSb$_2$O$_9$ has shown that despite the quantum origin of the 1/3 plateau phase in TLAF, its magnetic excitations can be described using semiclassical nonlinear spin-wave theory~\cite{kamiya2018nature}.

Thus, even though considerable progresses have been made in theoretical understanding of TLAF, good experimental realizations remain rare. CsYbSe$_2$ is one of the best known examples of TLAF without structural disorder~\cite{SI}, with negligible interlayer coupling and relatively small magnetic energy scale~\cite{Xing20192}, which allows us to tune the magnetic ground state by moderate magnetic field. In this Letter, using elastic and inelastic neutron scattering, heat capacity, and density-matrix renormalization group (DMRG) methods, we first demonstrate that CsYbSe$_2$ can be described by a {\it nearly isotropic} (Heisenberg) model. We then reveal the pronounced changes in the magnetic excitation spectrum as it evolves with magnetic field and associate these changes with the field-driven phase transitions and thus provide comprehensive overview of spin dynamics in TLAF Heisenberg model.

We prepared high-quality single crystals of CsYbSe$_2$ using a flux method according to the previous report~\cite{Xing20191}. The refinement of single-crystal X-ray diffraction shows no Cs/Yb site mixing in our crystals~\cite{SI}. The anisotropy of the $g$ factor was studied by electron spin resonance (ESR) measurements [Fig.~\ref{fig1}(b)], under the same conditions used in other Yb-based delafossites~\cite{Sichelschmidt_2019}. Heat capacity data with temperature down to $T=50$~mK and magnetic field up to $B=9$~T were collected with a semi-adiabatic compensated heat-pulse method in a dilution refrigerator~\cite{Kihara2013}.
For the INS measurements, about 200 millimeter-sized single crystals were co-aligned in the $(HHL)$ scattering plane on thin copper plates to get a mosaic sample with a mass about 0.5~g and mosaicity $\sim$ 2\degree~\cite{SI}. The neutron scattering experiments were performed at the time-of-flight Cold Neutron Chopper Spectrometer (CNCS)~\cite{CNCS1,CNCS2} at the Spallation Neutron Source, Oak Ridge National Laboratory. Measurements were carried out with an incident neutron energy $E_{\mathrm{i}}$ = 3.32 meV ($\lambda_{\mathrm{i}}$ = 4.96 {\AA}), providing an energy resolution of 0.11~meV [full width at half maximum (FWHM) at the elastic position]. The magnetic fields were applied along the vertical direction [$-K$~$K$~0] using a 5~T cryomagnet equipped with a dilution refrigerator to cool the sample down to $T$~=~70~mK. The sample was rotated along the vertical axis in a wide angle range (210$\degree$) to get a good coverage in both the energy and momentum space. We used the software packages~\textsc{MantidPlot}~\cite{Mantid} and~\textsc{Horace}~\cite{Horace} for data reduction and analysis.

Due to the localized character of 4$f$ states, the effective spin-spin interactions in CsYbSe$_2$ are mainly limited by the NN exchange.
The general NN anisotropic-exchange Hamiltonian constrained by the symmetries of the triangular lattice can be written as~\cite{li2015rare,Liyaodong2016,Liyaodong2018,zhu2018topography,maksimov2019anisotropic}
\begin{align}  \label{Hamiltonian}
	&\mathcal{H} = J\sum_{\langle i,j \rangle} ({S}^x_i{S}^x_j + {S}^y_i{S}^y_j + \Delta{S}^z_i{S}^z_j ) - \mu_B g_{ab} B \sum_i{S}^y_i \nonumber \\
	&+ \sum_{\langle i,j \rangle} 2J_{\pm \pm} [(S^x_iS^x_j - S^y_iS^y_j)\tilde{c}_{\alpha}
	- (S^x_iS^y_j + S^y_iS^x_j)\tilde{s}_{\alpha}]
	\nonumber \\
	&+ \sum_{\langle i,j \rangle} J_{z \pm} [(S^y_iS^z_j + S^z_iS^y_j)\tilde{c}_{\alpha}
	- (S^x_iS^z_j + S^z_iS^x_j)\tilde{s}_{\alpha}],
\end{align}
where $\langle i,j \rangle$ denotes NN sites, $0 \le \Delta \le 1$,  $\mu_B$ is the Bohr magneton, $g_{ab}$ is the in-plane $g$-tensor component, $B$ is the strength of the magnetic induction, which is parallel to the in-plane $\bf{y}$ direction [Fig.~\ref{fig1}(a)].
$\tilde{c}_{\alpha} = \cos{\tilde{\varphi}_{\alpha}}$ and $\tilde{s}_{\alpha} = \sin{\tilde{\varphi}_{\alpha}}$, in which bond angles $\tilde{\varphi}_{\alpha}$ are angles between the primitive vectors of the lattice $\bm\delta_\alpha$ and the $x$ axis [Fig.~\ref{fig1}(a)], $\tilde{\varphi}_{\alpha}=\{0,2\pi/3,-2\pi/3\}$.
The exchange interaction parameters follow, $J_{zy}=J_{yz}=J_{z\pm}$, $J_{zz}=\Delta J$, with $J=(J_{xx}+J_{yy})/2$ and $J_{\pm\pm}=(J_{xx}-J_{yy})/4$~\cite{maksimov2019anisotropic}.

Hamiltonian (\ref{Hamiltonian}) has been extensively explored to establish the phase diagram of the general NN TLAF model~\cite{Liyaodong2016,Luo2017, zhu2017disorder, zhu2018topography, Iaconis2018, maksimov2019anisotropic}.
It was found that the phase diagram is dominated by magnetically ordered single-$\mathbf{Q}$ phases, which are ferromagnetic, 120$^{\circ}$ N$\acute{e}e$l, dual 120$^{\circ}$, and two different stripe states \cite{maksimov2019anisotropic}.
For CsYbSe$_2$, from the fitting of the experimental magnetization curve and the width of the 1/3 magnetization plateau with grand canonical DMRG method, we found that the XXZ anisotropy is small, $i.e.$, $\Delta \approx 1$~(see details in~\cite{SI}). The best fitting parameters are $J$ = 0.544 meV, $g_{ab}$ = 4.148.
To examine the dynamical properties of the system~\eqref{Hamiltonian}, dynamical DMRG calculations~\cite{Jeckelmann,dynamics1D} on 61-site open clusters under open boundary conditions were performed (see details in~\cite{SI}).
Having in mind that (i) we did not observe the nonreciprocal spin-wave spectra which is caused by the presence of anisotropic bond-dependent interaction $J_{z\pm}$ ~\cite{maksimov2019anisotropic} (see details in~\cite{SI}), and (ii) $J_{z\pm}$ and $J_{\pm\pm}$ are expected to be small compared to $J$~\cite{Ranjith2019naybse}, we considered the bond-independent XXZ part of Hamiltonian~(\ref{Hamiltonian}) only, neglecting the bond-dependent anisotropic terms.

\begin{figure}[tb]
\center{\includegraphics[width=0.95\linewidth]{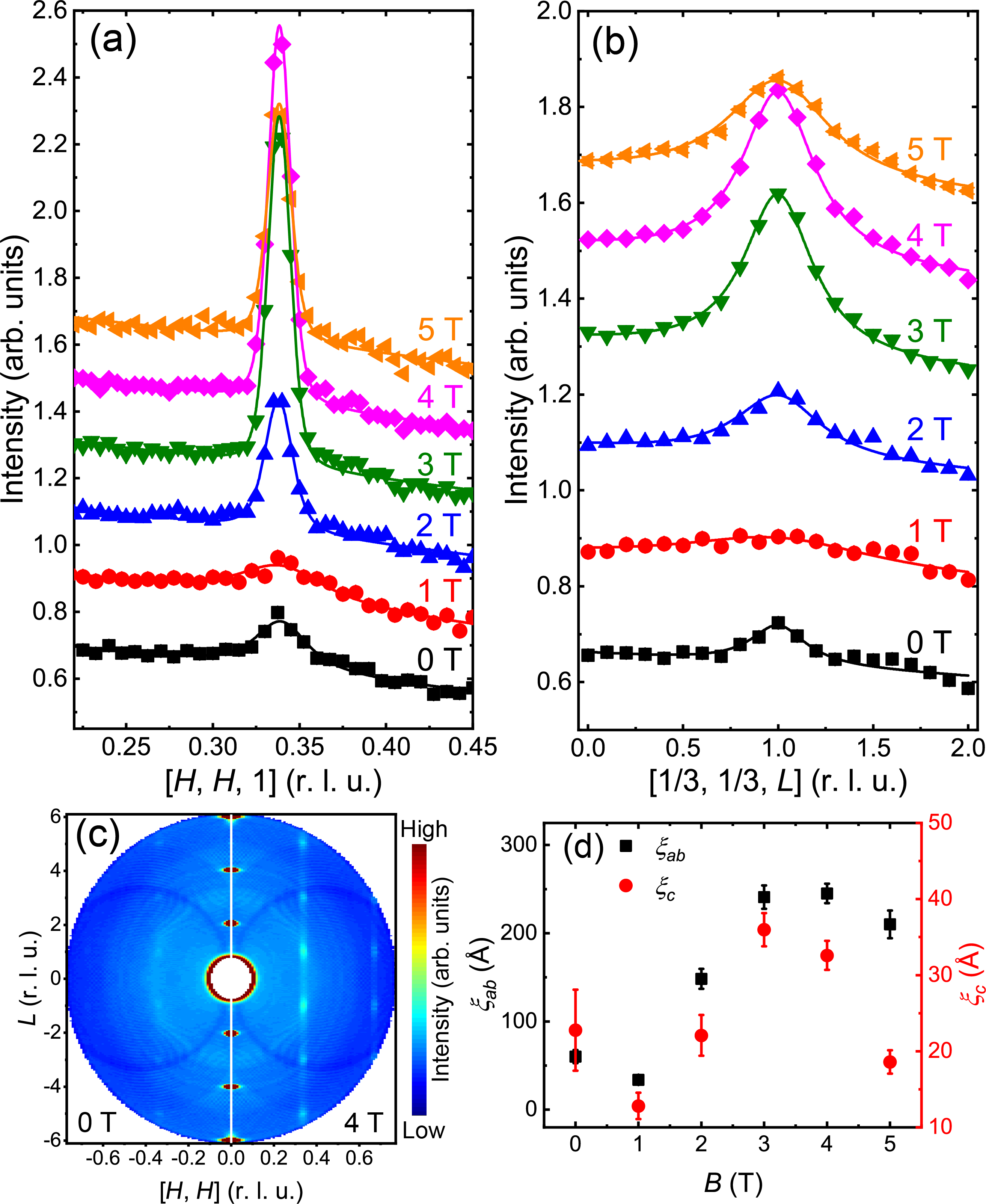}}
  \caption{~Elastic neutron scattering signals ($E$ = 0 $\pm$ 0.05) in $(HHL)$ plane at 70 mK. (a) 1D constant energy cuts along [$H$ $H$ 1]. (b) 1D constant energy cuts along [1/3 1/3 $L$]. The solid curves are fitting results of the Voigt function with linear backgrounds. The data points and fitting curves have been shifted vertically for clarity. (c) 2D zero-energy slices in $(HHL)$ plane for 0 T (left-half part) and 4 T (right-half part). In panels (a)--(c), the integrated range along [$-K$ $K$ 0] direction is $K$ = [$-$0.05, 0.05]. (d) The in-plane and out-of-plane correlation length obtained from the Fourier transform of the magnetic Bragg peak width at $Q = (1/3, 1/3, 1)$.
  }
  \label{Elastic}
\end{figure}

In order to reveal which phases are stabilized at base temperature in different magnetic fields, we first analyze the elastic neutron scattering signal (obtained as neutron energy transfer $E=\hbar \omega=E_{\mathrm{i}} - E_{\mathrm{f}} = 0 \pm 0.05$~meV) in complement to the heat capacity data. The heat capacity as a function of temperature at different in-plane magnetic fields are presented in Figs.~\ref{fig1}(c) and \ref{fig1}(d). For the data shown in Fig.~\ref{fig1}(c), sharp $\lambda$-shape peaks that correspond to phase transitions of long-range orders can be identified in 3~T -- 5~T. While for the curves measured at magnetic fields $B \le 2$ T, and $B \ge 6$~T, only broad peaks/humps that indicate short-range correlations/orders can be observed [Fig.~\ref{fig1}(d)].

Elastic neutron scattering at 70~mK shows that weak magnetic order has formed at 0~T [Figs.~\ref{Elastic}(a)--\ref{Elastic}(c)]. In the two-dimensional (2D) constant energy slice at 0~T [left half in Fig.~\ref{Elastic}(c)], we can see weak intensity appears at $Q$ = (1/3, 1/3, $L$) ($L$ = odd integers), which correspond to a propagation vector $\bf{q}$ = (1/3, 1/3, 1). Since this weak order appears at zero field, we attribute it to the well-known 120$^{\circ}$ ordered state for the Heisenberg $S$ = 1/2 TLAF. At 4~T, we can see clearly rod of magnetic intensity elongated along the $[1/3~1/3~L]$ direction and peaked at $Q$ = (1/3,~1/3,~1) and equivalent positions [right half of Fig.~\ref{Elastic}(c)]. To characterize the peaks quantitatively we made the one-dimensional (1D) cuts along $[H~H~1]$ and $[1/3~1/3~L]$ at $Q$ = (1/3,~1/3,~1) [Figs.~\ref{Elastic}(a) and~\ref{Elastic}(b)].  At 0~T, the weak peak is broad along both directions and application of 1~T field suppresses it further. Above 1 T,  the magnetic peak evolves to be sharper and stronger with the increasing magnetic field and reaches a maximal intensity at 4~T. Further field increase up to 5~T causes minor decrease of the intensity. The data points shown in Figs.~\ref{Elastic}(a) and \ref{Elastic}(b) can be well fitted by Voigt function with a linear background~\cite{SI}. In order to make the spin-spin correlation more intuitive, we convert the peak width at $Q$ = (1/3, 1/3, 1) to the spin-spin correlation length ($\xi$) in real space taking into account the instrumental resolution, and the results are shown in Fig.~\ref{Elastic}(d)~\cite{SI}. For all the measured fields, $\xi_c$ $\textless$ 36 {\AA} ($\sim$ 2.2$c$, $c$ is lattice constant), which indicates a short-range spin-spin correlation along the $c$ axis. The spin-spin correlation in $ab$ plane also starts from short-range at 0~T ($\xi_{ab} \approx$ 60~{\AA}), reaches a minimum at 1~T ($\xi_{ab} \approx$ 34~{\AA}) and then evolves to be long-range above 3~T with $\xi_{ab}$ $\approx$~245 {\AA} ($\sim$ 60$a$, $a$ is lattice constant). The evolution of $\xi_{ab}$ is consistent with the heat capacity results in Figs.~\ref{fig1}(c) and \ref{fig1}(d), where sharp $\lambda$-shape peaks can only be observed in the long-range ordered states at 3~T -- 5~T, and the broad humps at 0~T -- 2~T are related to formation of the short-range magnetic orders.

\begin{figure*}[tb]
\center{\includegraphics[width=1\linewidth]{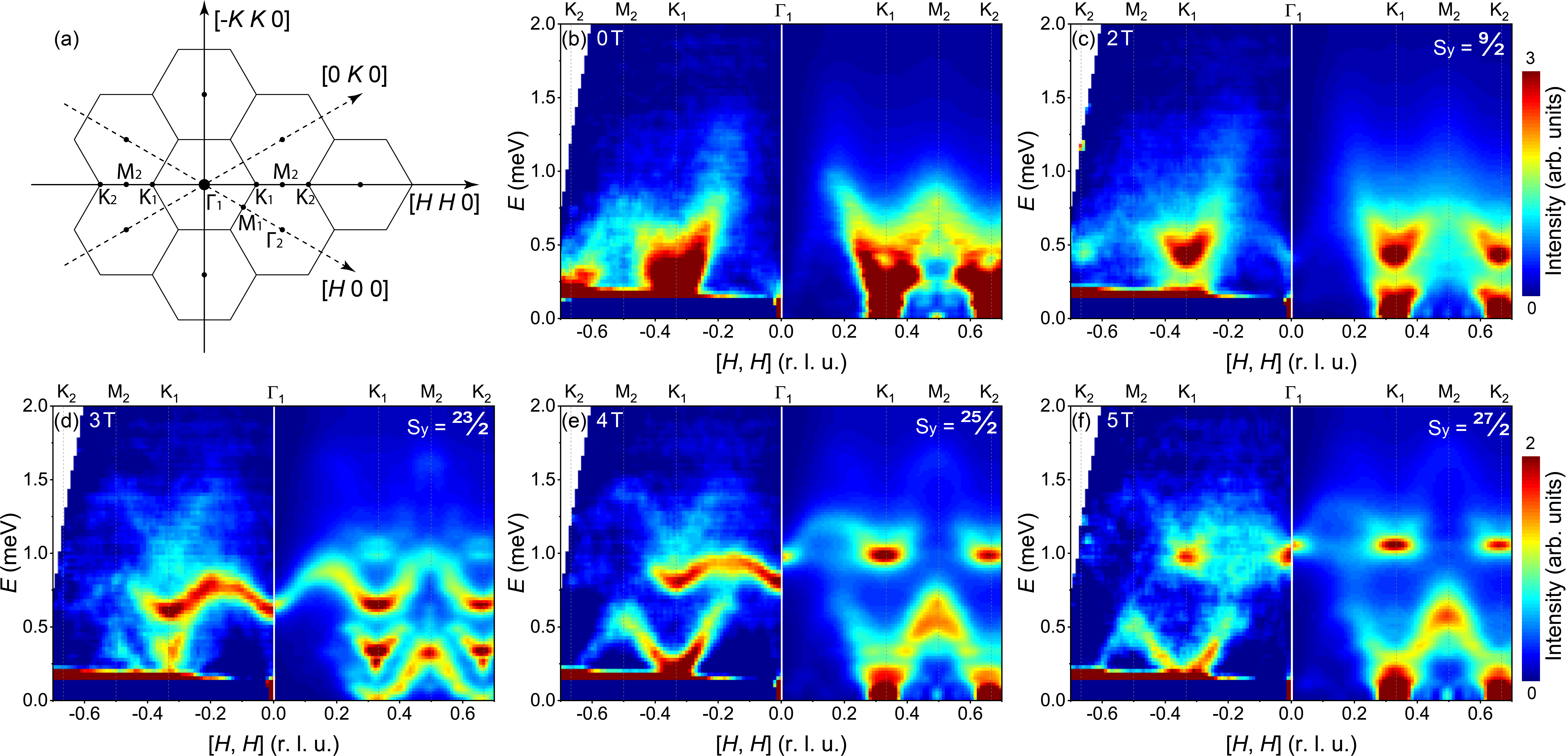}}
  \caption{~Spin excitation spectra along $[H~H]$ direction at different magnetic fields at 70~mK. (a) The schematic of the Brillouin zones in the $(H~K~0)$ plane. The high symmetry points and high symmetry directions are indicated by the dots and arrows, respectively. (b)--(f) Experimental (left-half panels) and calculated (right-half panels) dynamical spin structure factors $S(\mathbf{Q}, \omega)$ at different magnetic fields. All the experimental data shown here have been symmetrized according to the crystal symmetry. The integrated range along [$-K$ $K$] direction is $K$ = [$-$0.1, 0.1]. The $y$ components of the spin quantum numbers $S_y$ = 9/2, 23/2, 25/2, and 27/2 are used for indicating the strength of the magnetic fields in dynamical DMRG calculations~\cite{SI}.
  }
  \label{Dispersion}
\end{figure*}

According to our above analyses and the results in Ref.~\cite{Ranjith2019naybse}, we propose the following scenario for field-induced evolution of the magnetic structures in CsYbSe$_2$.  At base temperature, the system starts from a short-range 120$\degree$ magnetic order at 0~T, then gradually evolves to the oblique 120$\degree$ ($Y$-coplanar) phase near 2~T~\textless~$B$~\textless~3~T, and the collinear UUD order for $B$ $\ge$ 3~T. For higher field strength like 5~T and 6~T, the system is in a canted 2:1 coplanar state~\cite{Ranjith2019naybse,Garlea2019}. Combining the neutron diffraction results with the heat capacity data, we depict a schematic $B$-$T$ phase diagram [Fig.~\ref{fig1}(b)]. The solid and open circles in the phase diagram  correspond to the temperatures of the sharp peaks and broad humps in the heat capacity data shown in Figs.~\ref{fig1}(c) and \ref{fig1}(d), respectively.

We now turn to the field dependence of the dynamical properties of the TLAF  CsYbSe$_2$. Figure~\ref{Dispersion} summarizes the experimental spin excitation spectra (left-half panels) along the high symmetry direction [$H$ $H$] [the horizontal arrow in Fig.~\ref{Dispersion}(a)] at different magnetic fields in agreement with the corresponding DMRG calculations (right-half panels). Since the spin excitation has no intensity modulation and energy dispersion along [0 0 $L$] direction~\cite{SI}, we integrated a wide $L$ range (1.5~$\le$~$|\mathrm{L}|$~$\le$~5.5) to increase the statistics. The background of the spectra has been subtracted, and the details of the background subtraction can be found in supplemental material~\cite{SI}.

Similar to the previous reports~\cite{Xing20192,dai2020spinon},  the INS spectrum measured at 0~T shows diffusive continuum-like spin excitations accumulated around K point of the Brillouin zone [Fig.~\ref{Dispersion}(b)].
Except for the diffusive continuum-like excitation, the spectrum contains a branch of weak and diffusive spin excitation stems from the $\Gamma$ point. At 2~T, the spectrum opens a clear gap around the K point below $E$~$\sim$~0.4 meV [Fig.~\ref{Dispersion}(c)]. Below the gap energy, there is another low-energy mode that merges with the incoherent background. Further increasing the field to 3~T, the most impressive observation is that a strong and sharp spin-wave mode appears in the energy range 0.55 $\textless$ $E$ $\textless$ 0.8 meV (mode III) [Fig.~\ref{Dispersion}(d)]. Additionally, three other excitation branches can be identified. The lowest branch (mode I) with band top at $\sim$~0.4~meV is very weak and visible only in the momentum range $-$0.6 $\textless$ $H$ $\textless$ $-$0.4. The branch (mode II) between the mode I and the mode III concentrates its intensity around the K point and touches the mode III around 0.5~meV. The highest diffusive branch (mode IV) locates in the energy range 0.8 $\textless$ $E$ $\textless$ 1.5 meV. At 4 T, the mode I become very strong and the energy top is lifted to $\sim$ 0.7 meV, while the mode II disappears, and the mode III is lifted to 0.75 $\textless$ $E$ $\textless$ 1~meV. The mode IV keeps almost unchanged both in intensity and energy range. At the highest measured magnetic field, 5 T, the mode I becomes weaker than at 4~T. While, in the energy range 0.8 $\textless$ $E$ $\textless$ 1.2 meV, the mode III evolves to be a bow-tie-like continuum with intensity maximum located at $\Gamma$ and K point. The weakest mode IV seems to be still existent, but becomes weaker and merges with the bow-tie-like continuum.

 Previously, the spin dynamics in TLAF were described by different models, among which linear and non-linear spin wave theory (SWT) are the most conventional tools~\cite{Ma2016,ito2017,Mourigal2013,Zheng2006,Macdougal2021}. For instance, non-linear SWT was successfully applied to simulate the dispersion in TLAF Ba$_3$CoSb$_2$O$_9$ within the field-induced UUD phase~\cite{kamiya2018nature}. However, this model fails to reproduce the broadening magnon linewidth due to strong magnon-magnon interactions in the 120$\degree$ phase at zero field~\cite{Ma2016,ito2017}. Therefore, in this work we used a conceptually different approach and performed large-cluster DMRG method to calculate the spin dynamics explicitly at different fields (see supplemental material~\cite{SI} for details of DMRG calculations).
 The used exchange coupling parameter in the dynamical DMRG calculations is $J$ = 0.48 meV. The anisotropic components of $g$ factor ($g_{ab}$ = 3.25, $g_c$ = 0.3) extracted from the ESR measurements [Fig.~\ref{fig1}(b)] have been used in the calculations. As we show below, this simple model treated by dynamical DMRG provides surprisingly good qualitative account of the observed INS spectra.

The calculated dynamical spin structure factors, $S(\mathbf{Q}, \omega)$, are shown in Figs.~\ref{Dispersion}(b)--\ref{Dispersion}(f) along with the experimental spectra. For the low-field regime below the UUD phase as shown in Fig.~\ref{Dispersion}(b) and \ref{Dispersion}(c), our model well captures the overall behavior, including the strong, gapless, damped excitations at zero field and the two diffusive excitation modes with an energy gap between them at 2~T. At $B = 3$~T and 4~T, upon entering the collinear UUD phase, the calculated $S(\mathbf{Q}, \omega)$ not only reproduce the three stronger low-energy branches (modes I--III), but also gives reasonable description of the high-energy weak branch (mode IV) [Figs.~\ref{Dispersion}(c)--\ref{Dispersion}(d)]. It is worth noting that the calculated intensity of modes I--III mostly comes from the transverse (perpendicular to the direction of the magnetic field, $i.e.$, $\mathbf{y}$) structure factors $S^{xx}(\mathbf{Q}, \omega)$ and $S^{zz}(\mathbf{Q}, \omega)$, while the intensity of mode IV is mostly contributed by the longitudinal structure factor $S^{yy}(\mathbf{Q}, \omega)$ (see~supplemental information~\cite{SI} for different components of the calculated dynamical structure factors). This means that the modes I--III correspond to the single-magnon excitations, the weak and diffusive mode IV corresponds to the two-magnon continuum~\cite{kamiya2018nature,Chernyshev2009}. Above the UUD phase at $B = 5$~T, the low-energy mode, and the high-energy bow-tie-like continuum with resonance-like maximal intensity at $\Gamma$ and K points of the BZ are qualitatively reproduced by our DMRG calculation again [Fig.~\ref{Dispersion}(f)]. It should be noticed that the energy scale of the calculated spectra at some non-zero fields does not precisely match the experimental ones and we will discuss possible origins of the minor disagreements below.

The DMRG simulations of the Heisenberg TLAF model reproduce the major spectral features of the measured excitations:
(i) broad overdamped excitations at low-field 120$^{\circ}$ state [Fig.~\ref{Dispersion}(b)];
(ii) field-induced spin gap and the additional mode  stems from $\Gamma$ point [Fig.~\ref{Dispersion}(c)];
(iii) series of magnon modes and drastic increase of lifetime of magnons within the plateau phase [Figs.~\ref{Dispersion}(d) and \ref{Dispersion}(e)];
(iv) decay of magnon mode and resonance-like feature at K and $\Gamma$ points above the plateau phase [Fig.~\ref{Dispersion}(f)];
(v) multimagnon excitations at high energy [Figs.~\ref{Dispersion}(d)--\ref{Dispersion}(f)].
The consistency between experimental and numerical data indicates that the spin dynamics in CsYbSe$_2$ is primarily dictated by the isotropic exchange interaction. Therefore, our results provide comprehensive overview of field-induced spin dynamics in this perfect $S$ = 1/2 TLAF. The quantitative discrepancy between experimental data and the calculated spectra shown in Fig.~\ref{Dispersion} can be caused by the oversimplification of our model, which does not go beyond simple Heisinberg interaction as well as by finite-size effects in calculations. Another possible reason is that the strength of used magnetic fields in the calculations of $S(\mathbf{Q}, \omega)$ do not match all the experimental fields perfectly~\cite{SI}.

We further note that the zero-field spectrum shown in Fig.~\ref{Dispersion}(b) looks very similar to the continuum-like excitation observed in isostructural NaYbSe$_2$, which is claimed to be a QSL with a spinon Fermi surface~\cite{dai2020spinon}. As we have mentioned above, CsYbSe$_2$ is a perfect TLAF without disorder, the continuum-like excitation in Fig.~\ref{Dispersion}(b) cannot be a disorder-induced mimicry of QSL that was proposed in Ref.~\cite{zhu2017disorder}. Previous theoretical studies on TLAF have suggested that the QSL phase occurs at the border between 120$\degree$ and stripe phases, and the maximal extent of the QSL phase is achieved at the isotropic limit of the XXZ term~\cite{zhu2017disorder,zhu2018topography}. As the short-range 120$\degree$ magnetic order has been confirmed at zero field by elastic neutron scattering and the INS spectra has been reproduced by the DMRG calculations with an isotropic Heisenberg TLAF model, CsYbSe$_2$ may be located in a critical region between the QSL phase and 120$\degree$ phase at zero field. In this region the system is so unstable that even a weak perturbation of quantum fluctuations will push the system towards a QSL or magnetically ordered ground state.

In summary, we observed a short-range 120$\degree$ magnetic order at zero field in the $S$~=~1/2 TLAF CsYbSe$_2$. A schematic $B$-$T$ phase diagram was depicted on the basis of heat capacity and neutron diffraction results.  The coexistence of short-range 120$\degree$ magnetic order and highly damped continuum-like excitations at zero field suggests that CsYbSe$_2$ may be in a critical region between the QSL phase and 120$\degree$ phase. We report the first in-field single-crystal INS study on the family of Yb-based  dichalcogenide delafossites $A$Yb$Q_2$. The INS spectra evolve from the zero-field continuum-like excitations to relatively sharp spin wave modes in field-induced phases. The spin excitation spectra at different fields can be well described by an isotropic Heisenberg TLAF model with the dynamical DMRG method. This work provides a {\it nearly isotropic} NN Hamiltonian that is appropriate for the different quantum phases in an ideal $S$~=~1/2 TLAF compound, and sheds light on the understanding of the physics of TLAF compounds.

\section*{Acknowledgments}
We thank Colin McMillen for the help with the single-crystal X-ray diffraction refinement, U. Nitzsche for technical assistance, and B. Schmidt for the fruitful discussion. The research at the Oak Ridge National Laboratory (ORNL) is supported by the U.S. Department of Energy (DOE), Office of Science, Basic Energy Sciences (BES), Materials Sciences and Engineering Division.
This research used resources at Spallation Neutron Source, a DOE Office of Science User Facility operated by ORNL.
X-ray Laue alignment was conducted at the Center for Nanophase Materials Sciences (CNMS) (CNMS2019-R18) at ORNL, which is a DOE Office of Science User Facility.
S.E.N. acknowledges funding from the European Unions Horizon 2020 research and innovation program under the Marie Sk{\l}odowska-Curie grant agreement No 884104. S.N. acknowledges support by the Deutsche Forschungsgemeinschaft through SFB 1143 project no. A05.

This paper has been authored by UT-Battelle, LLC under Contract No. DE-AC05-00OR22725 with the U.S. Department of Energy. The United States Government retains and the publisher, by accepting the article for publication, acknowledges that the United States Government retains a non-exclusive, paid-up, irrevocable, world-wide license to publish or reproduce the published form of this manuscript, or allow others to do so, for United States Government purposes. The Department of Energy will provide public access to these results of federally sponsored research in accordance with the DOE Public Access Plan (http://energy.gov/downloads/doe-public-access-plan).
\bibliography{csybse2}

\end{document}


\title{Supplementary Materials: Field-Induced Spin Excitations in the Spin-1/2 Triangular-Lattice Antiferromagnet CsYbSe$_2$}
\author{Tao Xie}
\thanks{These authors contributed equally to this work}
\affiliation{Neutron Scattering Division, Oak Ridge National Laboratory, Oak Ridge, TN 37831, USA}
\author{Jie Xing}
\thanks{These authors contributed equally to this work}
\affiliation{Materials Science and Technology Division, Oak Ridge National Laboratory, Oak Ridge, Tennessee 37831, USA}
\author{S.~E.~Nikitin}
\affiliation{Paul Scherrer Institut, CH-5232 Villigen, Switzerland}
\author{S.~Nishimoto}
\affiliation{Department of Physics, Technical University Dresden, 01069 Dresden, Germany}
\affiliation{Institute for Theoretical Solid State Physics, IFW Dresden, 01069 Dresden, Germany}
\author{M.~Brando}
\affiliation{Max Planck Institute for Chemical Physics of Solids, N\"{o}thnitzer Str. 40, D-01187 Dresden, Germany}
\author{P.~Khanenko}
\affiliation{Max Planck Institute for Chemical Physics of Solids, N\"{o}thnitzer Str. 40, D-01187 Dresden, Germany}
\author{J.~Sichelschmidt}
\affiliation{Max Planck Institute for Chemical Physics of Solids, N\"{o}thnitzer Str. 40, D-01187 Dresden, Germany}
\author{L.~D.~Sanjeewa}
\affiliation{Materials Science and Technology Division, Oak Ridge National Laboratory, Oak Ridge, Tennessee 37831, USA}
\author{Athena~S.~Sefat}
\affiliation{Materials Science and Technology Division, Oak Ridge National Laboratory, Oak Ridge, Tennessee 37831, USA}
\author{A.~Podlesnyak}
\affiliation{Neutron Scattering Division, Oak Ridge National Laboratory, Oak Ridge, TN 37831, USA}
\maketitle

\begin{center}
{\bf A. Single-Crystal X-Ray Diffraction}
\end{center}

\begin{figure}[h]
\renewcommand\thefigure{S1}
\center{\includegraphics[width=1\linewidth]{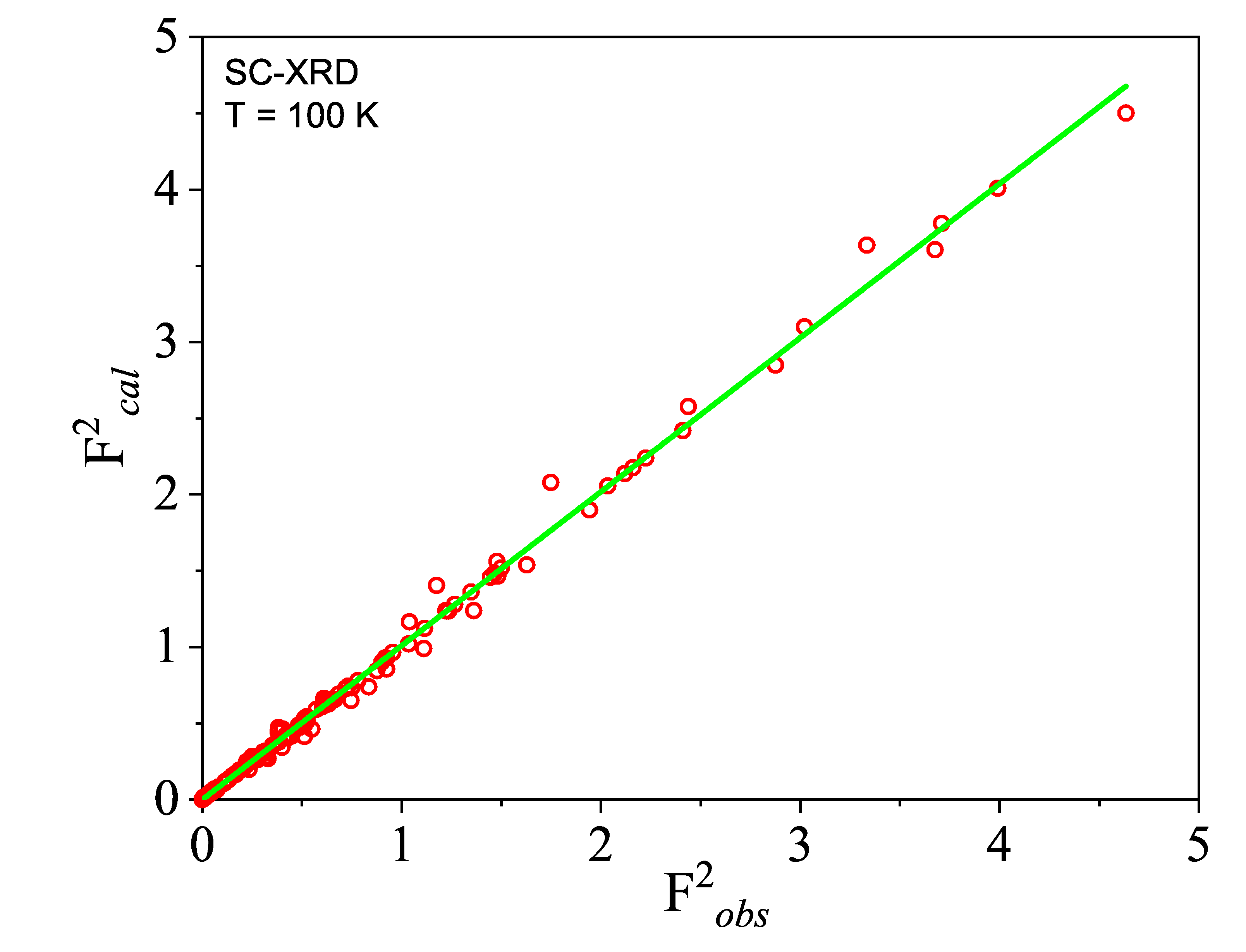}}
  \caption{~The Rietveld refinement results of the single-crystal X-ray diffraction data at 100 K. F$^2_{\mathrm{cal}}$ and F$^2_{\mathrm{obs}}$ are the calculated and observed structure factors, respectively.}
  \label{SCXRD}
\end{figure}

We analyzed the lattice structure carefully using single-crystal X-ray diffraction. The single-crystal X-ray diffraction experiments were performed on several batches of single crystals using Bruker Quest D8 single-crystal X-ray diffractometer. The structure was refined with Rietveld method using FullProf suite software package~\cite{FullProf}. We obtained very good structure solutions and didn't observe any hint of the mixed sites in our crystals. One of the crystallographic data determined by  single-crystal X-ray diffraction is presented in Table~\ref{tabSCXRD}, and one of the Rietved refinement results is plotted in Fig.~\ref{SCXRD}. For every refinement, the anisotropic displacement parameters and site occupancies were  refined as free variables, and no site disorder is observed. There is also no significant residual electron density, which suggests a stuffed lattice. All these results confirm the high quality of our single crystals. Additionally, we have reported the structure in the CCDC (1952075)~\cite{CCDC1952075}.

 Recently, we became aware of ref.\cite{dai2020spinon}, in which~ $4.8\% \pm 1\%$ of the Na/Yb site mixing was given by the single-crystal X-ray diffraction analysis on NaYbSe$_2$. It is important to mention that sodium ions tend to mix with the rare-earth ions due to the similar size of the ionic radii.  In fact, this kind of site mixing is very common in Na-based rare-earth orthophosphates, $e.g.$, Na$_{3(1+x)}$Yb$_{(2-x)}$(PO$_4$)$_3$~\cite{SALMON197985}.

\begin{table}[tb]
\caption{Crystallographic data of CsYbSe$_2$ determined by single-crystal X-ray diffraction. }
\begin{tabular}{|l|l|}
\hline
empirical formula                           & CsYbSe$_2$            \\ \hline
formula weight (g/mol)                      & 463.87             \\ \hline
$T$, K                                        & 100                \\ \hline
Crystal habitat                             & red plates         \\ \hline
Crystal dimensions, mm                      & $0.14\times0.10\times0.03$ \\ \hline
crystal system                              & hexagonal          \\ \hline
space group                                 & P63/mmc (No.194)   \\ \hline
\emph{a}, \AA                                        & 4.1466(2)          \\ \hline
\emph{c}, \AA                                        & 16.5050(1)         \\ \hline
volume, \AA$^3$                                  & 245.77(3)          \\ \hline
Z                                           & 2                  \\ \hline
\emph{D} (calc), g/cm$^3$                            & 6.268              \\ \hline
$\mu(\mathrm{Mo K_\alpha})$, mm$^{-1}$                             & 40.932             \\ \hline
\emph{F}(000)                                      & 386                \\ \hline
\emph{T}max, \emph{T}min                                  & 0.4415, 1.0000     \\ \hline
$\theta~~\mathrm{range}$                                     & 2.47-29.93         \\ \hline
reflections collected                       & 3895               \\ \hline
data/restraints/parameters                  & 171/0/9            \\ \hline
final $R~{[}I\textgreater   2\sigma(I){]} R_1, R_{w2}$ & 0.0274/0.0833      \\ \hline
final $R$ (all data) $R_1, R_{w2}$                  & 0.0284/0.0843      \\ \hline
GoF                                         & 1.343              \\ \hline
largest diff. peak/hole, e/\AA$^3~~~~$              & 1.405/-1.752       \\ \hline
\end{tabular}
\label{tabSCXRD}
\end{table}

\begin{center}
{\bf B. Elastic Neutron Scattering}
\end{center}

\begin{figure}[b]
\renewcommand\thefigure{S2}
\center{\includegraphics[width=1\linewidth]{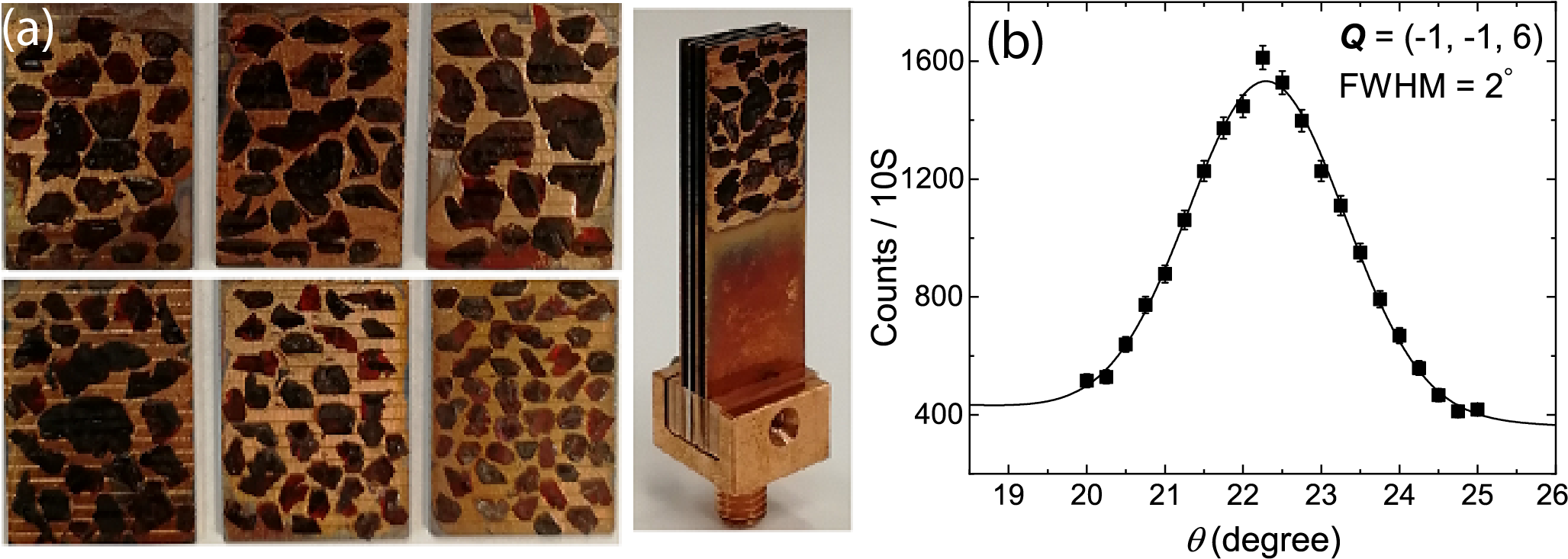}}
  \caption{~(a) Photos of the coaligned and assembled CsYbSe$_2$ crystals. (b) Rocking curve of the coaligned sample measured by neutron diffraction experiment. The solid line is Gauss fitting.
  }
  \label{sample}
\end{figure}

We coaligned about 0.5 g single crystals [about 200 pieces, see Fig.~\ref{sample}(a)] on copper plates to get a mosaic sample. The rocking curve from neutron diffraction measurement shows that the mosaicity ($\sim$ 2\degree) of our sample is pretty good [Fig.~\ref{sample}(b)].

\begin{figure*}[b]
\renewcommand\thefigure{S3}
\center{\includegraphics[width=1\linewidth]{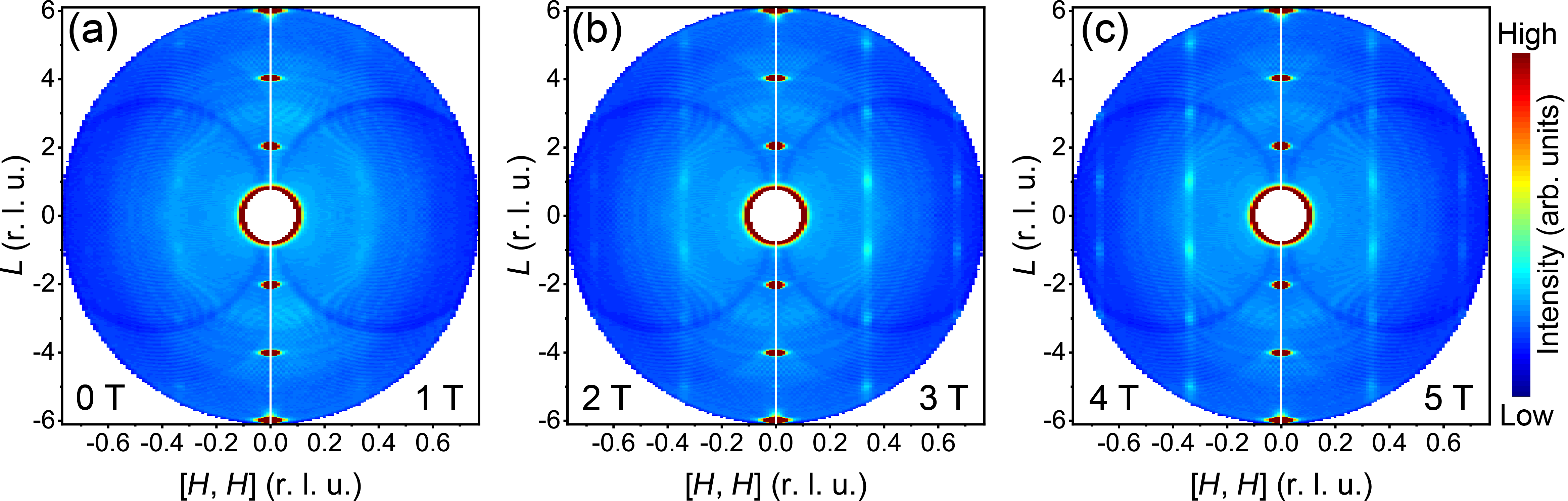}}
  \caption{~2D constant energy slices at zero energy under different magnetic fields. The data was symmetrized about the [$H, H$, 0] axis.
  }
  \label{2D_elastic}
\end{figure*}

Figure~\ref{2D_elastic} summarizes the 2D zero-energy slices of our neutron scattering results at 0 T $-$ 5 T. At 0~T, we can see weak but clear intensity appears at $Q$ = (1/3, 1/3, $L$) ($L$ = odd integers), while these peaks become weaker and almost invisible at 1 T. For $B$~$\ge$~2 T,  rods of magnetic intensity elongated along the $[1/3~1/3~L]$ direction and peaked at $Q$ = (1/3,~1/3,~$L$) ($L$ = odd integers) can be observed. The data has been symmetrized about the [$H~H$ 0] axis. The strong spots at $Q$ = (0, 0, $L$) ($L$ = even integers) are nuclear peaks.

\begin{figure*}[tb]
\renewcommand\thefigure{S4}
\center{\includegraphics[width=0.7\linewidth]{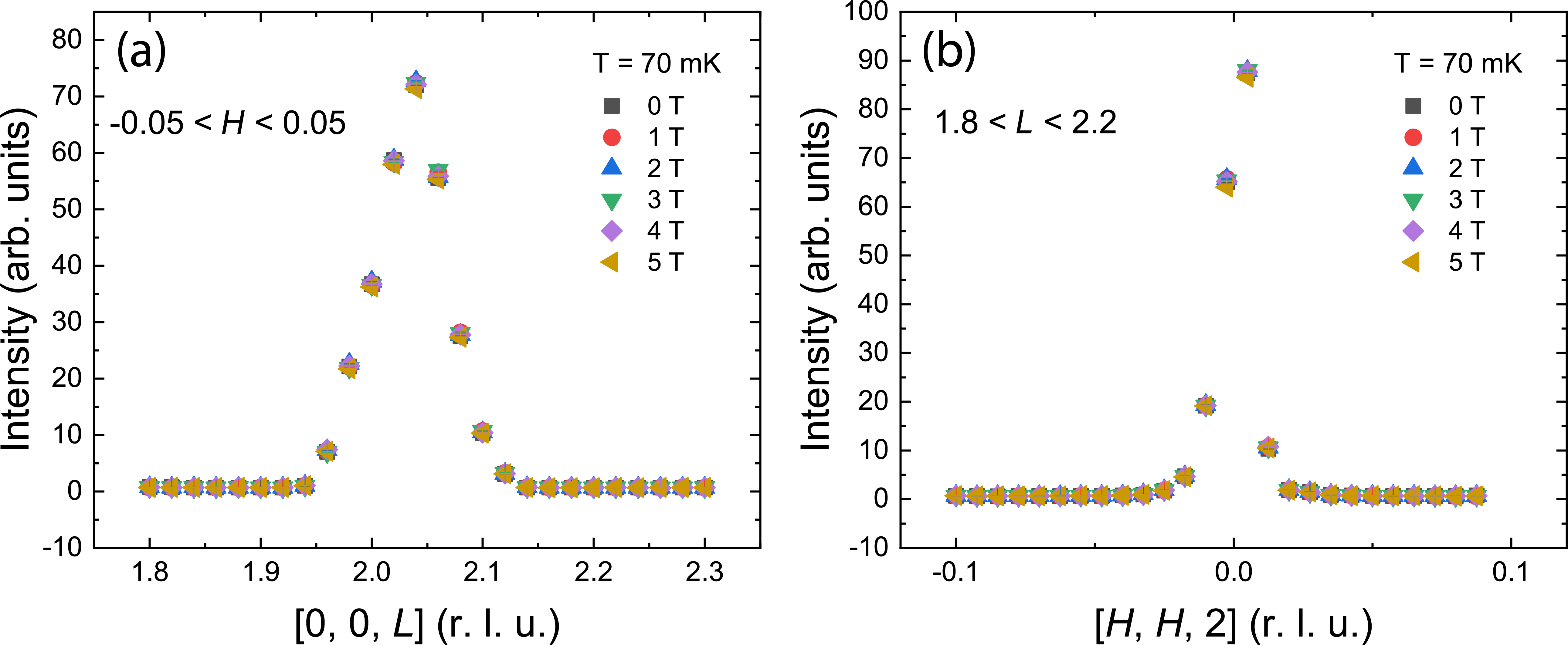}}
  \caption{~1D zero energy cuts along the [0, 0, $L$] and [$H, H$, 2] direction at nuclear peak (0, 0, 2) for different magnetic fields.
  }
  \label{nuclear}
\end{figure*}

\begin{figure*}[tb]
\renewcommand\thefigure{S5}
\center{\includegraphics[width=1\linewidth]{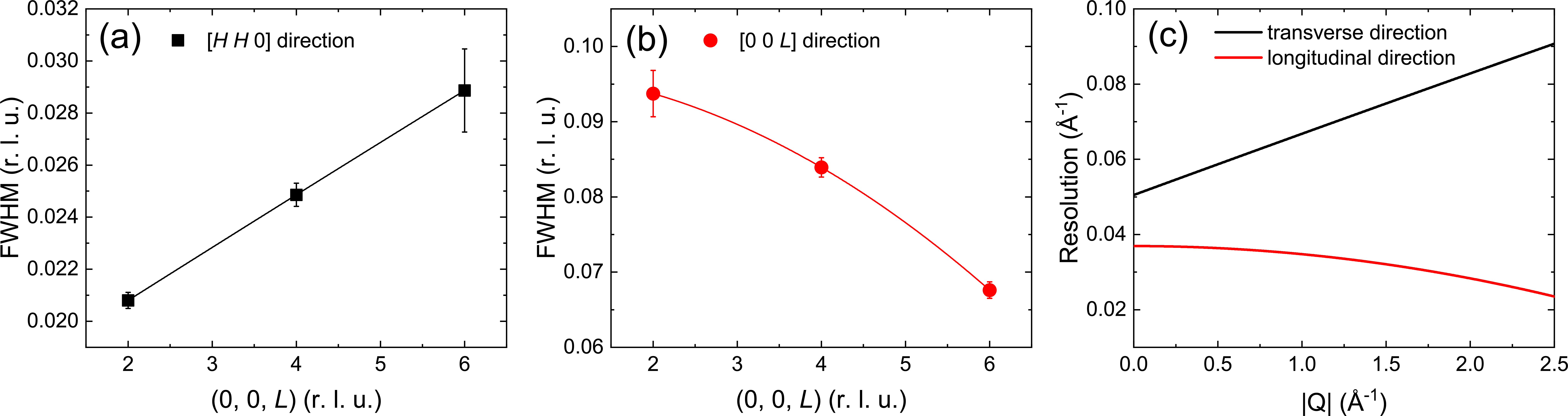}}
  \caption{~Instrumental momentum resolution estimation. (a) The FWHM of the nuclear peaks cut along [$H~H$ 0] direction at $Q$ = (0, 0, 2), (0, 0, 4), (0, 0, 6). (b) The FWHM of the nuclear peaks cut along [0, 0, $L$] direction at $Q$ = (0, 0, 2), (0, 0, 4), (0, 0, 6). The solid lines are polynomial fittings. (c) The estimated instrumental $Q$-resolution. The transverse resolution was estimated from the FWHM in panel (a), the longitudinal resolution was estimated from the FWHM in panel (b).
  }
  \label{resolution}
\end{figure*}

For a specific magnetic Bragg peak, the spin-spin correlation length $\xi$ can be estimated with formula $\xi$ = 2$\pi$/$\sqrt{w^2-R^2}$~\cite{Young2013},where $w$ is the full width at half maximum (FWHM) of the magnetic peak, $R$ represents the instrumental momentum resolution at the magnetic peak. The estimation of the instrumental resolution is simply described as following. We first made 1D cuts along the [$H$~$H$~0] (transverse direction) and [0~0~$L$] (longitudinal direction) directions at nuclear peaks (002) (Fig.~\ref{nuclear}), (004), and (006), then performed Gaussian fitting on the 1D peaks. The FWHMs of these nuclear peaks reflect the momentum resolution of the instrument at positions $Q$ = (0, 0, 2), (0, 0, 4), and (0, 0, 6). We then plotted the FWHM as a function of $L$, and performed polynomial fittings on FWHM [Figs.~\ref{resolution}(a) and~\ref{resolution}(b)]. At last, we approximately estimated the general resolution curves as a function of $\left|\bf{Q}\right|$ along the transverse and longitudinal directions for a random momentum transfer $\bf{Q}$ [Fig.~\ref{resolution}(c)].

To obtain the FWHM of the magnetic peaks, we performed Voigt function fittings on the 1D cuts of the magnetic peaks shown in Figs.~2(a) and 2(b) in main text. The Voigt function, convolution of a Gauss function and a Lorentz function, is defined as following:
\begin{align}
y&=y_{0}+f_{1}(x)\cdot f_{2}(x)
\end{align}
where $f_{1}(x)=\frac{2A}{\pi}\frac{w_{L}}{{4(x-x_{c}})^2+w_{L}^2}$ is the Lorentz function with $w_{L}$ as its FWHM and $x_c$ as the peak center, $f_{2}(x)=\sqrt{\frac{4\ln 2}{\pi}}\frac{e^{-\frac{4\ln 2}{w_{G}^2}x^2}}{w_{G}}$ is the Gauss function with its peak center fixed at $x$ = 0, and peak area equal 1. $w_{G}$ is the FWHM of the Gauss function, which should be fixed to the instrumental resolution $R(\left|\bf{Q}\right|)$ for a specific $\bf{Q}$. The FWHM of the Voigt function $w_V$ = $0.5346\cdot w_{L}$ +  $\sqrt{0.2166\cdot w_{L}^2+w_{G}^2}$. Having the $w_V$ and instrumental resolution $R(\left|\bf{Q}\right|)$, we can estimate the correlation length $\xi$ shown in Fig.~2(d) with formula $\xi$ = 2$\pi$/$\sqrt{w_{V}^2-R^2}$~\cite{Young2013}. The estimated $\xi$ at $Q$ = (1/3, 1/3, 1) for different fields are shown in Fig. 2(d) in the main text.

\begin{center}
{\bf C. 2D Excitation}
\end{center}

\begin{figure*}[tb]
\renewcommand\thefigure{S6}
\center{\includegraphics[width=1\linewidth]{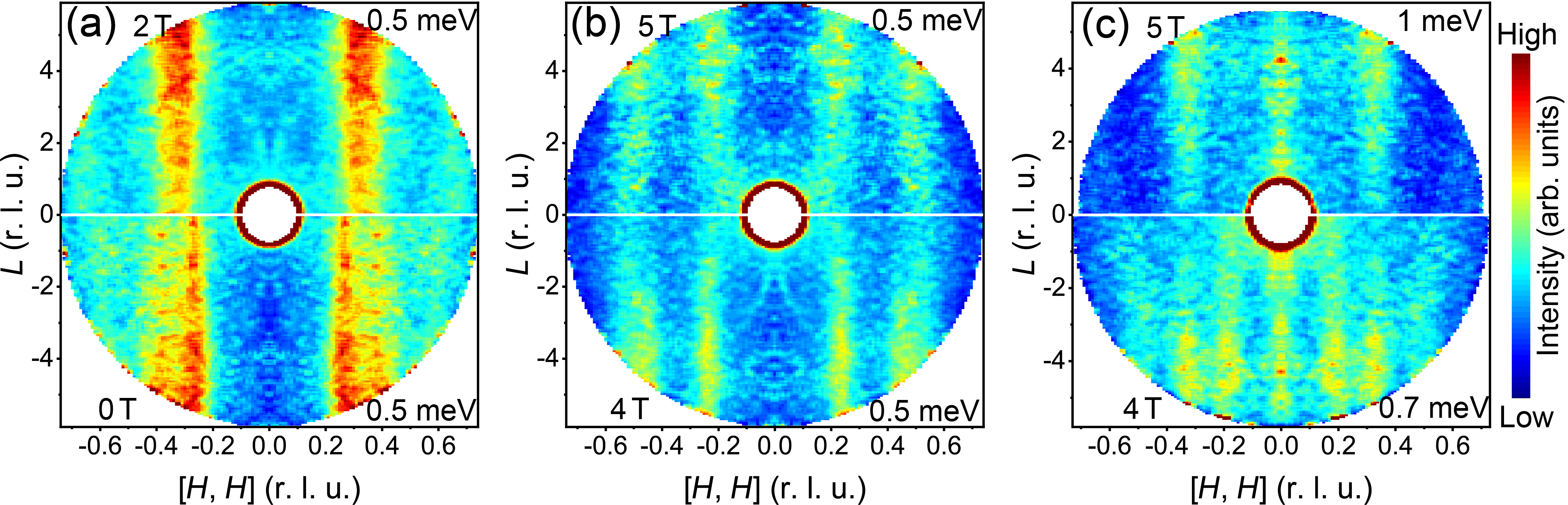}}
  \caption{~2D constant energy slices at different energies and magnetic fields. The data was symmetrized about the [0, 0, $L$] axis.
  }
  \label{2D_inelastic}
\end{figure*}

Consistent with previous reports~\cite{Xing20192,dai2020spinon}, the spin excitations in our measurements keep highly 2D in the whole energy range under all the measured magnetic fields. Several typical 2D constant-energy slices are presented in Fig.~\ref{2D_inelastic}, the intensity looks like several rods along the $L$ direction, which is a sign of 2D spin excitations. Thus, we can integrate the data in $L$ direction as much as possible when we discuss the spin dynamics in other directions, and we have integrated a wide $L$ range with 1.5~$\le$~$|\mathrm{L}|$~$\le$~5.5 for the data shown in Fig. 3 of the main text.

\begin{figure*}[tb]
\renewcommand\thefigure{S7}
\center{\includegraphics[width=1\linewidth]{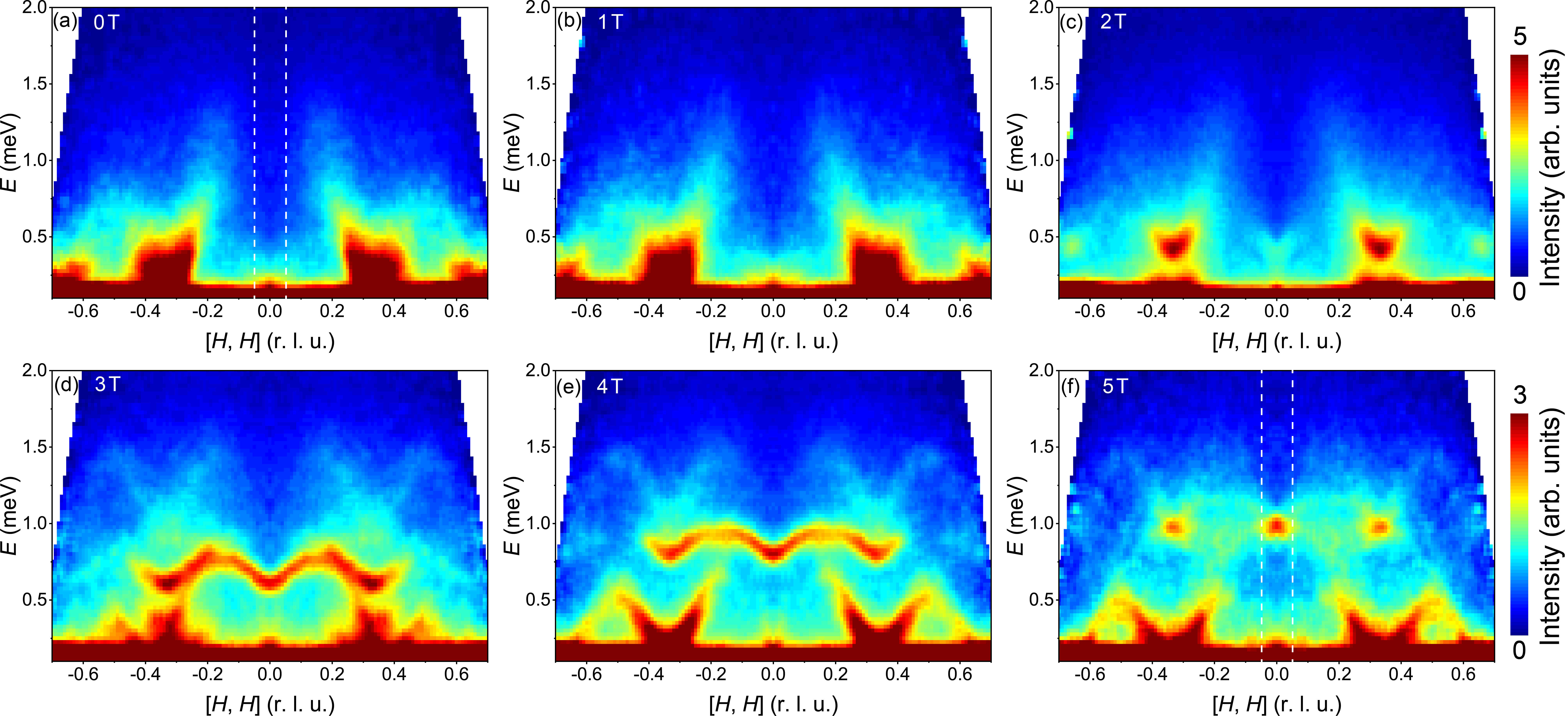}}
  \caption{~Raw data of the Spin excitation spectra along $[H~H]$ direction at different magnetic fields at 70 mK. The region between the vertical white dashed lines in panel (a) represents for the background used for background subtraction in Fig.~\ref{SI_Dispersion2}, while the region between the vertical white dashed lines in panel (f) represents for the region where we extract the background used for background subtraction in Fig.~\ref{SI_Dispersion3}. The shown data has been symmetrized.
  }
  \label{SI_Dispersion1}
\end{figure*}

\begin{figure*}[tb]
\renewcommand\thefigure{S8}
\center{\includegraphics[width=1\linewidth]{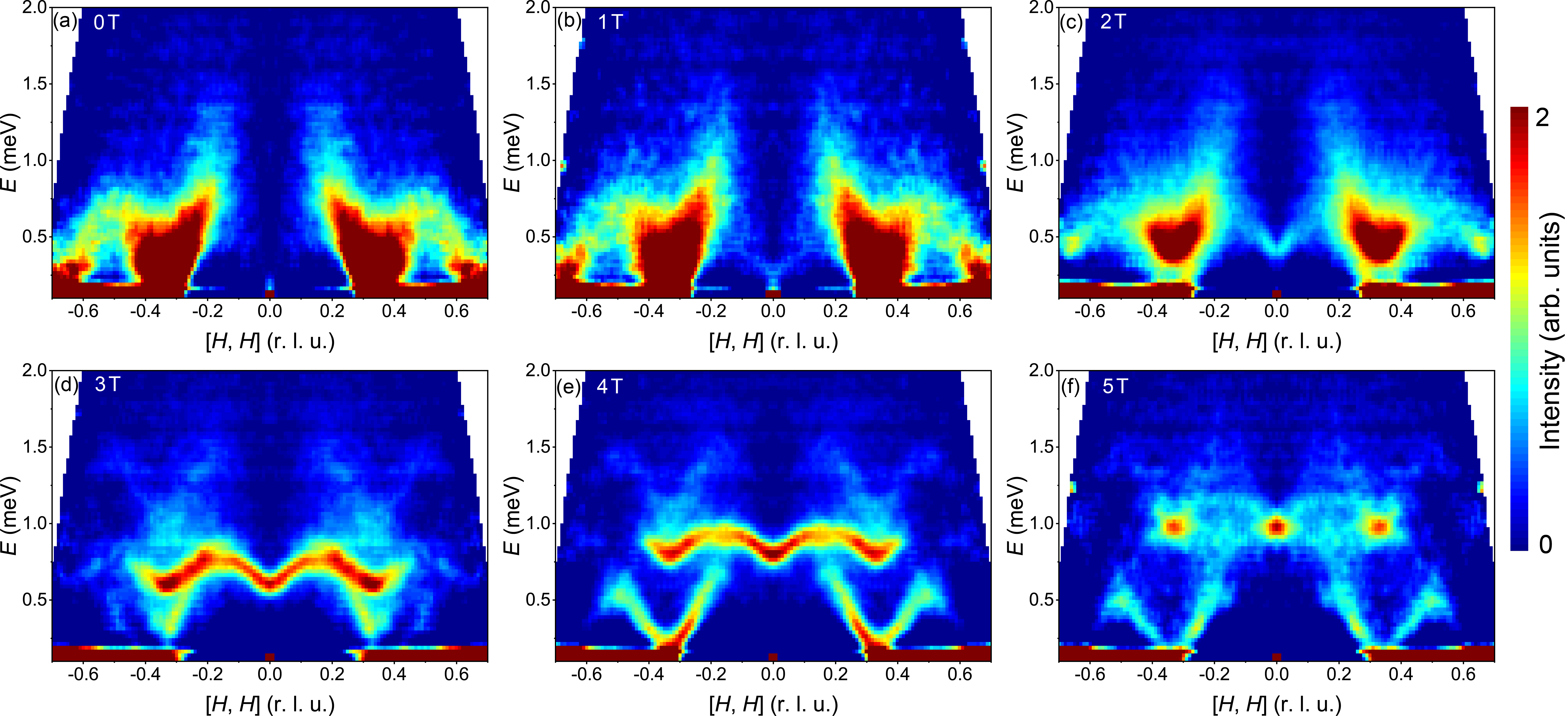}}
  \caption{~Spin excitation spectra along $[H~H]$ direction at different magnetic fields at 70 mK. The background was subtracted using the data between the vertical white dashed lines in Fig.~\ref{SI_Dispersion1}(a). The shown data has been symmetrized.
  }
  \label{SI_Dispersion2}
\end{figure*}

\begin{figure*}[tb]
\renewcommand\thefigure{S9}
\center{\includegraphics[width=1\linewidth]{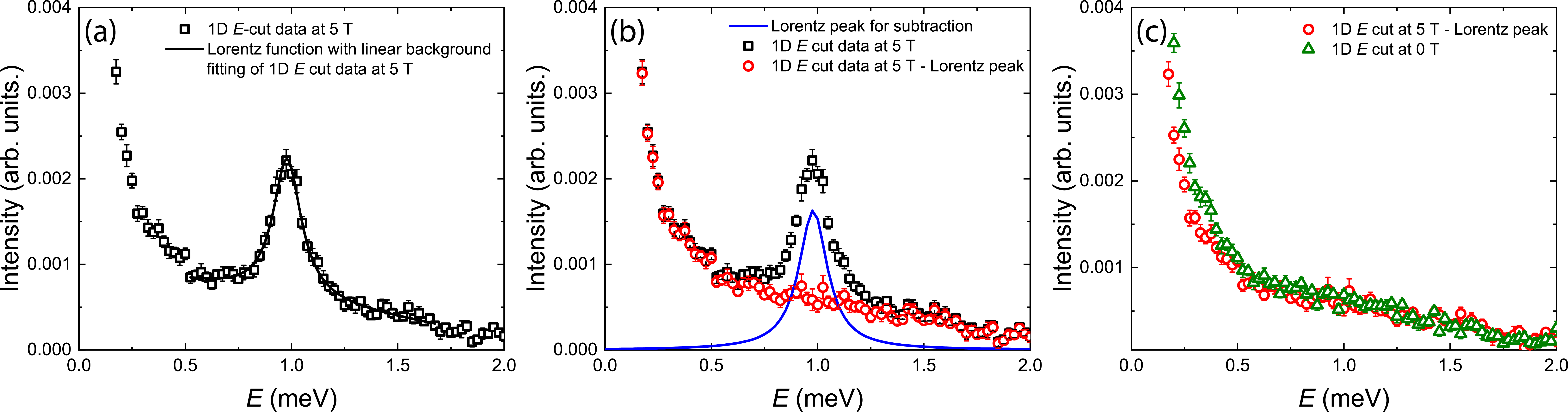}}
  \caption{~The definition of the subtracted background in the INS spectra in Fig. 3. (a) The 1D constant-$Q$ cut with $Q$ = (0~$\pm$~0.05, 0~$\pm$~0.05, 0) [the $Q$ region between the vertical white dashed line in Fig.~\ref{SI_Dispersion1}(f)] at 5 T, the black line is the Lorentz fitting with a linear background. (b) The black data points are the same with that shown in panel (a), the blue line is the fitting Lorentz peak in panel (a) after the linear background subtracted. Subtracting the blue line from the black data points subtract will give the the red data points, which will be treated as the real background for background subtraction in Fig. 3. (c) Comparison between the 1D constant-$Q$ cut with $Q$ = (0~$\pm$~0.05, 0~$\pm$~0.05, 0) at 0 T and the red data points shown in panel (b), which are the backgrounds used for background subtraction in Figs.~\ref{SI_Dispersion2} and \ref{SI_Dispersion3}, respectively.
  }
  \label{BKG}
\end{figure*}

\begin{figure*}[tb]
\renewcommand\thefigure{S10}
\center{\includegraphics[width=1\linewidth]{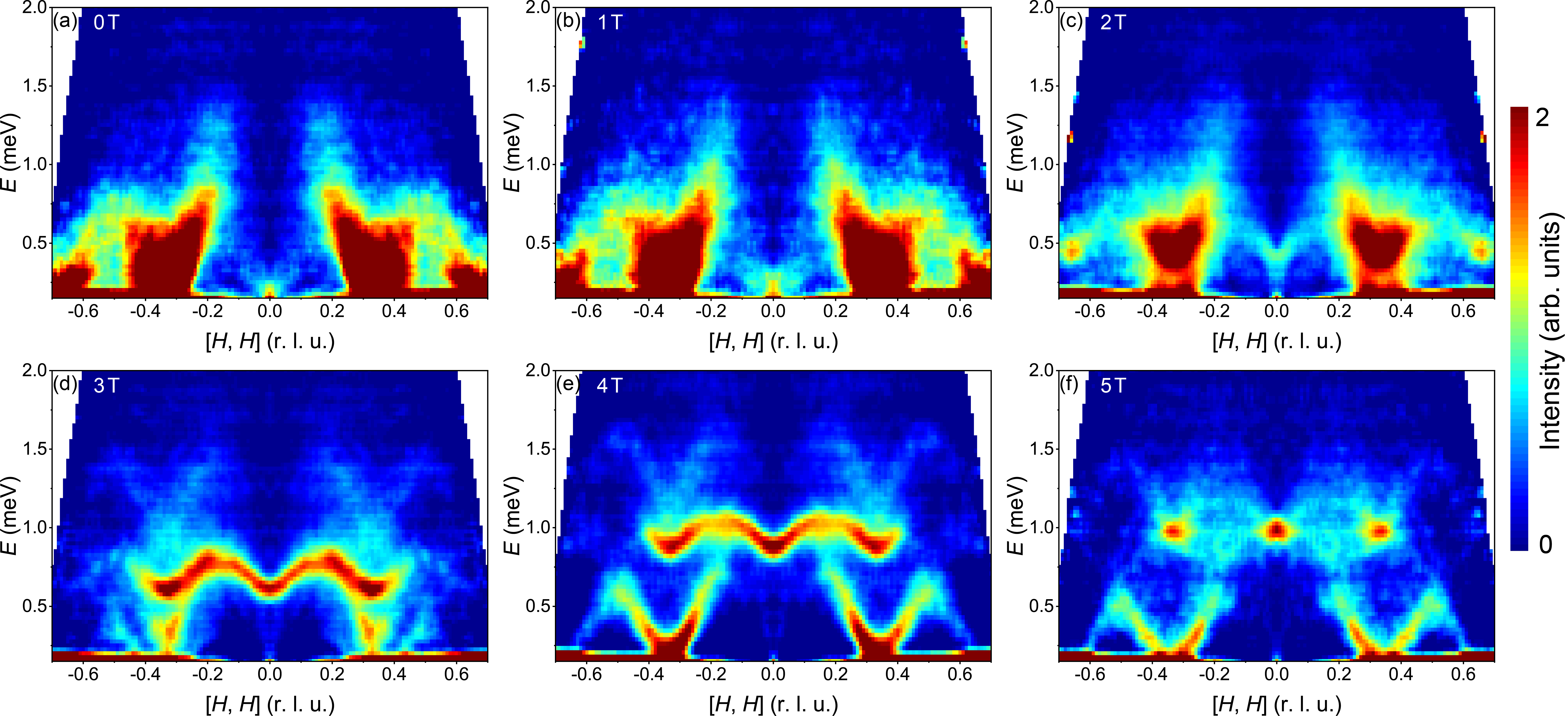}}
  \caption{~Spin excitation spectra along $[H~H]$ direction at different magnetic fields at 70 mK. The background was subtracted using the background extracted from the the data between the vertical white dashed lines in Fig.~\ref{SI_Dispersion1}(f), $i.e.$, using the red data points in Fig.~\ref{BKG}(b) as background. The shown data has been symmetrized.
  }
  \label{SI_Dispersion3}
\end{figure*}

\begin{center}
{\bf D. Background Definition and Subtraction for INS Spectra}
\end{center}

The INS spectrum is always contaminated by background (BG) contribution due to incoherent neutron scattering as well as scattering from the sample environment. Raw data of the spin excitation spectra along $[H~H]$ direction at different magnetic fields at 70 mK are shown in Fig.~\ref{SI_Dispersion1}. Since there is no obvious magnetic excitation intensity at $\Gamma$ point in the spectrum of 0 T shown in Fig.~\ref{SI_Dispersion1}(a), a very natural idea is to use the intensity around $\Gamma$ point at 0 T as the BG [The region between the vertical white dashed lines in Fig.~\ref{SI_Dispersion1}(a)] for BG subtraction. The BG subtraction results are presented in Fig.~\ref{SI_Dispersion2}. Though the subtraction in Fig.~\ref{SI_Dispersion2} looks not bad, we noted that there are some oversubtraction for energy range below 0.5 meV. This means that there is weak magnetic excitation intensity in the region between the vertical white dashed lines in Fig.~\ref{SI_Dispersion1}(a), which makes it not a perfect BG.

To subtract the BG better, we make use of the fact that the spectrum at $\Gamma$ point, $I(\mathbf{Q} = \mathbf{0}, E)$, measured at 5~T consists of a single and well-defined inelastic peak, which can be well described by the Lorentz function as shown in Fig.~\ref{BKG}(a). Assuming that within the energy window, 0.5~meV~$\le$~$E$~$\le$~1.5~meV, the BG signal can be approximated by a linear function, we fitted the inelastic intensity using the following equation:
\begin{align}
    I(E) = a_0 + a_1E +  \frac{I_0 W_0^2}{ (E-E_0)^2 + W_0^2}
    \label{eq:BG}
\end{align}
where $I_0$, $W_0$, and $E_0$ characterize the intensity, peak width, and peak center of the inelastic peak, respectively. Then the fitted peak [shown by the blue line in Fig.~\ref{BKG}(b)] was subtracted from the raw spectrum and the remaining intensity was considered as the BG. The raw spectrum, the fitted inelastic peak, and the remaining intensity after the subtraction of the Lorentz peak are shown in Fig.~\ref{BKG}(b). In Fig.~\ref{BKG}(c), we can see that the spectrum at 5 T with the Lorentz peak subtracted has a little weaker intensity than the raw spectrum at 0 T below 0.5 meV. This means that using the BG extracted from 5 T spectrum around $\Gamma$ point will avoid the oversubtraction to some extent.

We further assumed that the BG has no $\mathbf{Q}$ dependence and subtracted the obtained BG from the INS data at all reciprocal space $\mathbf{Q}$. The BG subtracted spectra at $B = 0~-~5$~T are shown in Figs.~\ref{SI_Dispersion3}.

\begin{center}
{\bf E. Absence of Non-Reciprocal Behavior}
\end{center}
\begin{figure}[tb]
\renewcommand\thefigure{S11}
\center{\includegraphics[width=.9\linewidth]{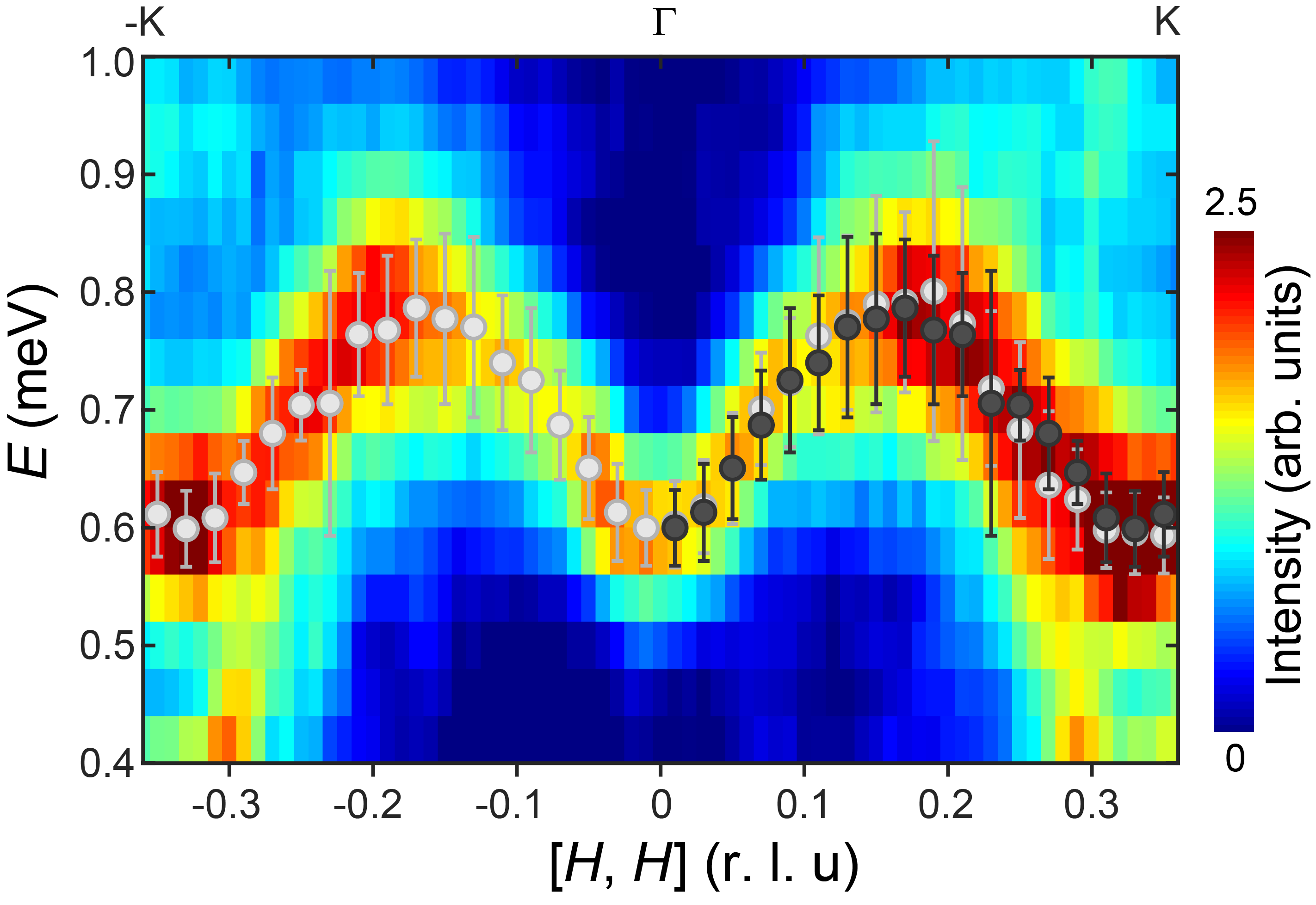}}
  \caption{~Magnon dispersion measured along $-$K$~\rightarrow~\Gamma~\rightarrow$~K direction of the reciprocal space measured at $B = 3$~T. The light grey points show fitted positions of the magnon mode. The dark grey points show mirror reflection of the negative part of the dispersion curve $\varepsilon$($-$K~$\rightarrow\Gamma$) towards positive part, $\varepsilon$($\Gamma\rightarrow$~K).
  }
  \label{SI_nonreciprocal}
\end{figure}

Authors of recent theoretical work~\cite{maksimov2019anisotropic} made an intriguing prediction that if the $J_{z\pm}$ term is present in Hamiltonian (1) in the main text, it will break the inversion symmetry. As the consequence, the magnon spectrum becomes nonreciprocal with $\varepsilon(\mathbf{Q}) \neq\ \varepsilon(\mathbf{-Q})$. To check whether such a behavior is realized in CsYbSe$_2$ we studied a sharp magnon mode measured at 3~T using non-symmetrized dataset, see Fig.~\ref{SI_nonreciprocal}.
For that, we fitted individual constant-$\mathbf{Q}$ cuts along the chosen part in reciprocal space and the resulted positions are shown by light grey dots in Fig.~\ref{SI_nonreciprocal}. To check whether the magnons are reciprocal we reflected the left part of the dispersion, $\varepsilon$($-$K~$\rightarrow\Gamma$), towards the right part, $\varepsilon$($\Gamma\rightarrow$~K) and the resulting data are shown by dark grey dots in Fig.~\ref{SI_nonreciprocal}. One can see that within the error bar of least square fittings both curves coincides with each other.
We further quantified the agreement between $\varepsilon(\mathbf{Q})$ and $\varepsilon(\mathbf{-Q})$ by calculating the $\chi^2$ factor:
\begin{align}
    \chi^2 = \frac{1}{N} \sum_{i=1,..,N} \frac{(E(\mathbf{Q}_i) - E(-\mathbf{Q}_i))^2}{dE^2(\mathbf{Q}_i) + dE^2(-\mathbf{Q}_i)}.
\end{align}
The obtained $\chi^2 = 0.364$ indicates very high quantitative match, meaning that $\varepsilon(\mathbf{Q}) = \varepsilon(\mathbf{-Q})$ and thus we can exclude the presence of a valuable $J_{z\pm}$ in CsYbSe$_2$.

\begin{center}
{\bf F. Fitting of Experimental Magnetization Curve}
\end{center}

\begin{figure}[t]
\renewcommand\thefigure{S12}
\centering
\includegraphics[width=1\linewidth]{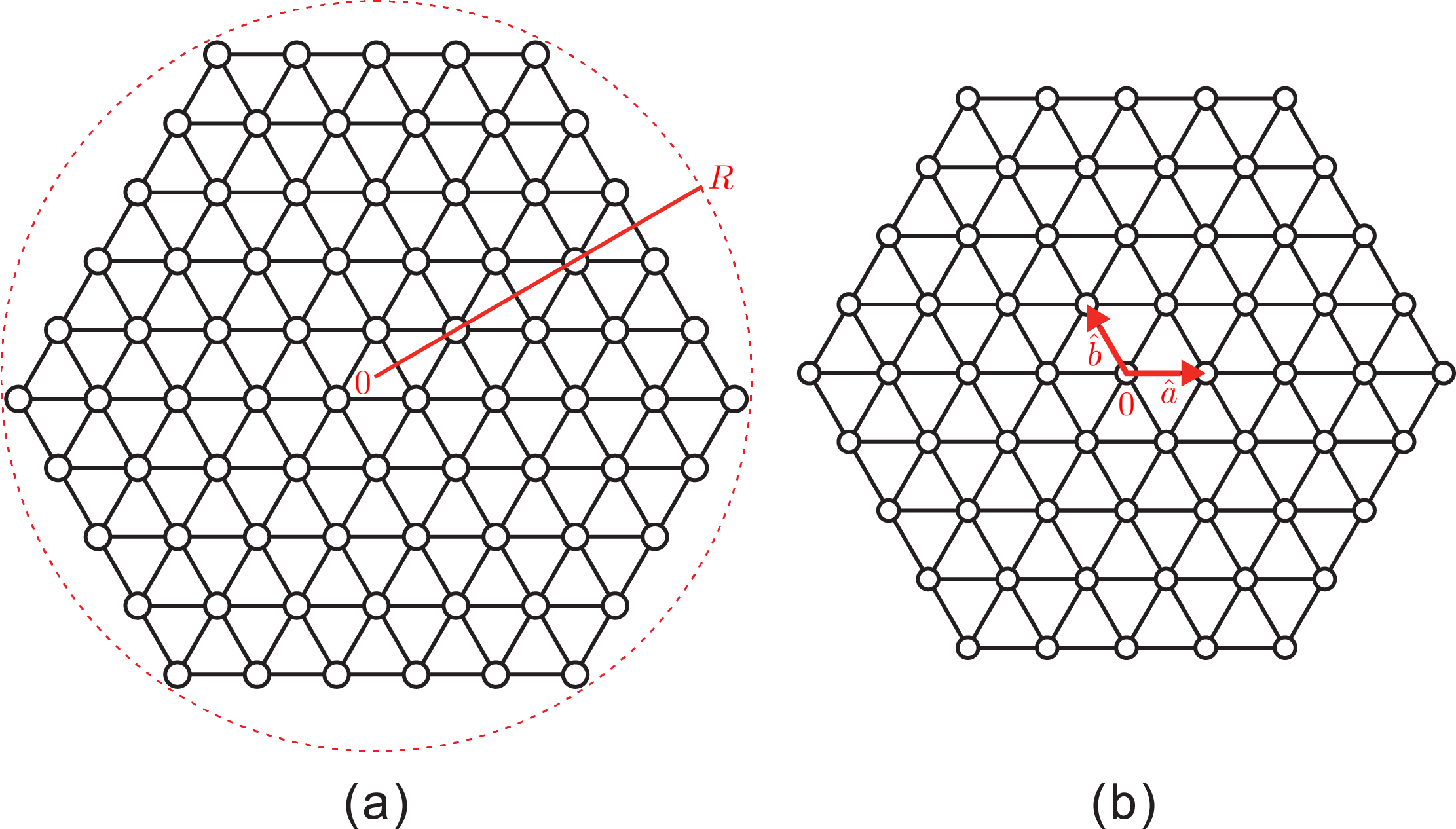}
\caption{
(a) $75$-site open cluster used in the ground canonical DMRG calculations.
(b) $61$-site open cluster used in the dynamical DMRG calculations.
}
\label{N61cluster}
\end{figure}

Here we perform a fitting of the experimental magnetization curve. To calculate the magnetization curve numerically, we apply the grand canonical DMRG method~\cite{Hotta2013}. The technique gives the infinitesimally small response to the change in the external field. Thus, the physical quantities we get mimic their thermodynamic limit within the order of $10^{-3}$ for two dimensions. The method smoothly divides the finite size cluster into the centre part and the edges, and the main part reproduces the continuous bulk response by using the nearly zero-energy edge state as a buffer. In a system with fixed size and shape, we introduce the modulation of the energy scale by an externally given function,
\begin{align}
f(r)= \frac{1}{2}\left[1+\cos\Big( \frac{\pi r}{R}\Big)\right],
\label{polarf}
\end{align}
at a location $r$, which smoothly deforms the Hamiltonian from the maximum
at the centre of the system ($r=0$) to zero energy at the open cluster
edges ($r=R$). After obtaining the proper eigen wave function of the deformed Hamiltonian, we evaluate the magnetization $M=\langle S^y(r \sim 0) \rangle$ (the direction of applied field is along $y$ axis). This is because the system optimizes the wave function so as to realize the centre value, $\langle S^y(r \sim 0) \rangle$, to its thermodynamic limit at a given magnetic field. As we need a result of a bulk system, equivalently spanned along three different directions of a triangular lattice, we choose a hexagonal cluster with 75 sites for the present calculation [see Fig.~\ref{N61cluster}(a)].

\begin{figure}[t]
\renewcommand\thefigure{S13}
\center{\includegraphics[width=1\linewidth]{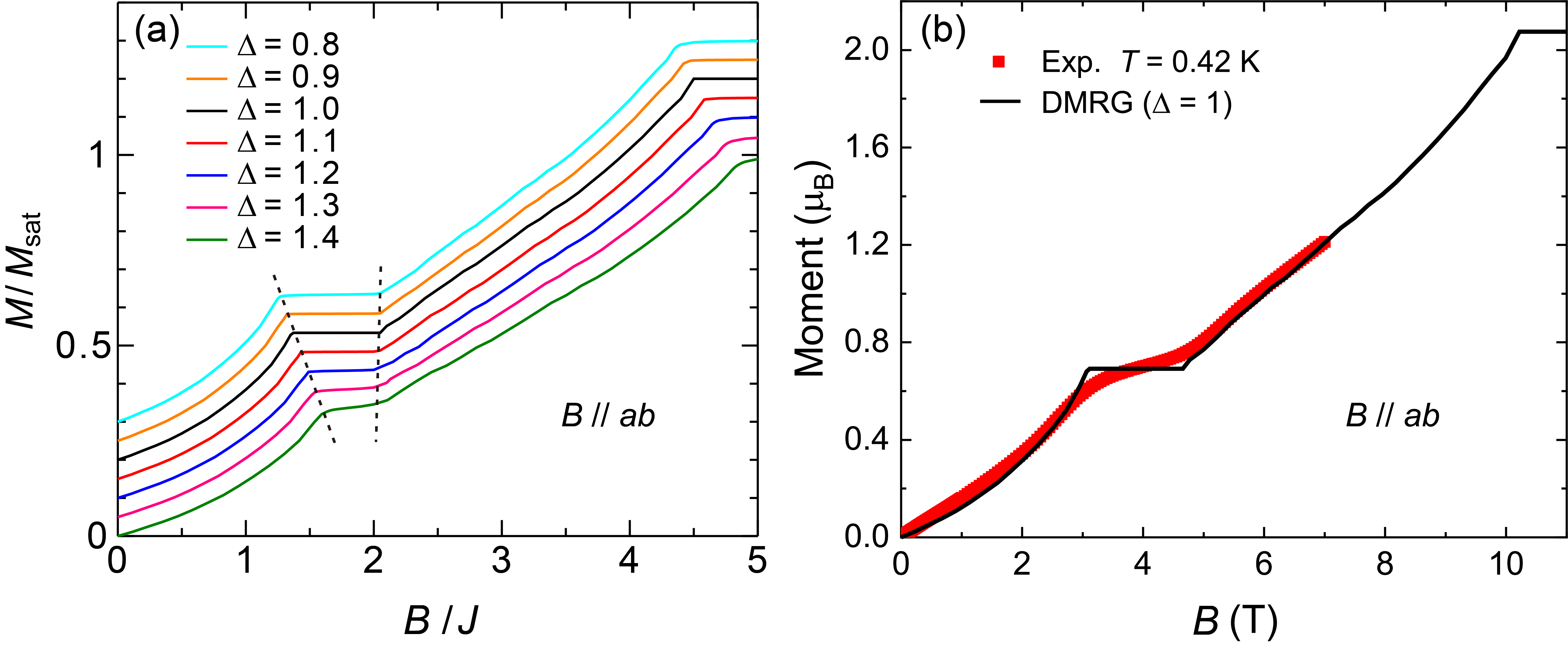}}
  \caption{~(a) DMRG calculation of magnetization curves of NN exchange interactions with different XXZ anisotropy. The curves have been shifted vertically by 0.05 for clarity. The dotted lines are guides to eyes to indicate the widths of the plateaus. (b) Fitting of the low-temperature experimental magnetization curve for $B$ $\parallel$ $ab$ by the ground canonical DMRG with an isotropic NN Heisenberg model. The parameters used for the fitting are $J$ = 0.544 meV, $g_{ab}$ = 4.148.
  }
  \label{SI_MH}
\end{figure}

The calculations were performed with a nearest-neighbor (NN) XXZ Hamiltonian
\begin{align}  \label{Hamiltonian_SI}
	&\mathcal{H} = J\sum_{\langle i,j \rangle} ({S}^x_i{S}^x_j + {S}^y_i{S}^y_j + \Delta{S}^z_i{S}^z_j ) \nonumber \\
	&- \mu_B g_{ab} B \sum_i{S}^y_i. 
\end{align}
The XXZ anisotropy ($\Delta$) dependence of calculated magnetization curves are illustrated in Fig.~\ref{SI_MH}(a). These magnetization curves are obtained by using conventional and grand canonical DMRG methods in a complementary style (The saturation field and the width of plateau are estimated by conventional DMRG method because the finite-size effect is known to be small for these quantities). We find that the width of the 1/3 magnetization plateau is sensitive to $\Delta$ and the plateau remains nearly flat at $\Delta=0.8-1.4$~[Fig.~\ref{SI_MH}(a)] although the total $S_y$ is no longer conserved except at $\Delta=1$. Therefore, a relatively unique fitting would be expected. The fitting result using the grand canonical DMRG method is shown in Fig.~\ref{SI_MH}(b). A best fitting
is obtained by setting $J=0.544$ meV, $g_{\rm ab}=4.148$, and assuming nearly isotropic exchange coupling ($\Delta \approx 1$).

\begin{center}
{\bf G. Calculation of Dynamical Structure Factor}
\end{center}

In the main text, we calculated the dynamical spin structure factor of
triangular Heisenberg model using dynamical DMRG technique~\cite{Jeckelmann}. While the ground state of triangular
Heisenberg model is a 120$^\circ$ antiferromagnetic state, a small spatial
anisotropy may lead to the other collinear or dimer orders~\cite{Starykh2007}. Accordingly, it is strongly expected to use a spatially isotropic cluster for the calculations. In this paper, we thus use a $N=61$ open cluster shown in Fig.~\ref{N61cluster}(b). The application of open boundary conditions enables us to obtain the physical quantities quite accurately by DMRG. Another advantage to use open boundary conditions is that the periodicity of wave function is not restricted; namely, the periodicity of magnetic structure can be continuously varied with changing model parameters as well as with applying external field, $etc$.

The dynamical spin structure factor is defined as
\begin{eqnarray}
\nonumber
S^{\gamma\bar{\gamma}}(\mathbf{Q},\omega) &=& \frac{1}{\pi}{\rm Im} \langle \psi_0 | (S^\gamma_{\mathbf{Q}})^\dagger \frac{1}{\hat{H}+\omega-E_0-{\rm i}\eta} S^\gamma_{\mathbf{Q}} | \psi_0 \rangle\\
&=& \sum_\nu |\langle \psi_\nu |S^\gamma_{\mathbf{Q}}| \psi_0 \rangle|^2 \delta(\omega-E_\nu+E_0),
\label{spec}
\end{eqnarray}
where $\gamma$ is $z$ or $-(+)$, $| \psi_\nu \rangle$ and $E_\nu$ are the $\nu$-th eingenstate and the eigenenergy  of the system, respectively ($\nu=0$ corresponds to the ground state). Since the momentum is ill-defined under
open boundary conditions, we usually use the eigenstates of the particle-in-a-box
problem to define the momentum-dependent spin operators
\begin{align}
S^\gamma_{\mathbf{Q}} = \sqrt{\frac{2}{N+1}} \sum_l \sin(2\mathbf{Q} \cdot \mathbf{R}) S^\gamma_{\mathbf{R}},
\label{operator_obc}
\end{align}
with quasi-momentum $Q_{(a,b)}=\pi z_{(a,b)}/(L_{(a,b)}+1)$ for integers
$1 < z_{(a,b)} \le L_{(a,b)}$ and $\mathbf{R}=(a,b)$. However, it is known that this
definition of $S^\gamma_{\mathbf{Q}}$ does not provide a quantitative agreement
of $S^{\gamma\bar{\gamma}}(\mathbf{Q},\omega)$ to that in the thermodynamic
limit at some momenta~\cite{dynamics1D}. This problem can be resolved
by defining the spin operators as
\begin{align}
S^\gamma_{\mathbf{Q}} = \sqrt{\frac{1}{N}} \sum_l e^{i2{\mathbf{Q}}\cdot{\mathbf{R}}} S^\gamma_{\mathbf{R}},
\label{operator_imp}
\end{align}
where the momentum $Q_{(a,b)}$ can take any value between $0$ and $\pi/2$
because of the unrestricted period of ground state wave function under open
boundary conditions. These spin operators were used to obtain the spectra
shown in this paper. The 120$^\circ$ antiferromagnetic order corresponds to a propagation vector
$\mathbf{Q}=(H,H)\pi=(1/3,1/3)\pi$. In fact, the spectrum calculated with
Eq.~(\ref{operator_imp}) is equivalent to the averaged ones obtained with
Eq.~(\ref{operator_obc}) and with the following spin operators
\begin{equation}
S^\gamma_{\mathbf{Q}} = \sqrt{\frac{2}{N+1}} \sum_l \cos(2\mathbf{Q} \cdot \mathbf{R}) S^\gamma_{\mathbf{R}}.
\label{operator_cos}
\end{equation}

The dynamical DMRG approach is based on a variational principle so that we
have to prepare a `good trial function' of the ground state. Therefore, we
keep $\sim 3000$ density-matrix eigenstates to obtain the ground state in the first ten DMRG sweeps and then keep $\sim 800$ density-matrix eigenstates to calculate the excitation spectrum. In this way, the maximum discarded weight is about $10^{-5} - 10^{-4}$.

\begin{figure*}[tb]
\renewcommand\thefigure{S14}
\center{\includegraphics[width=1\linewidth]{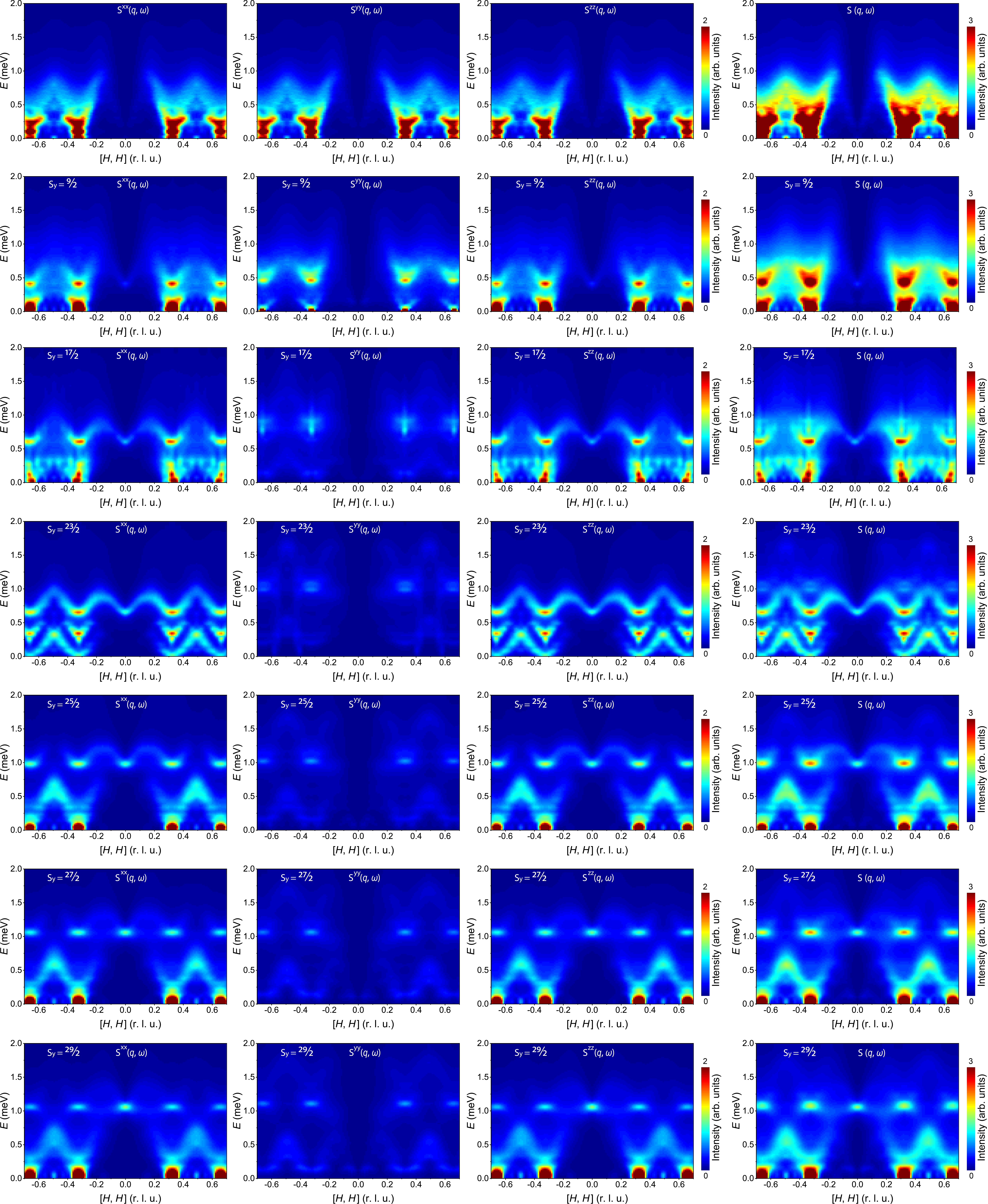}}
  \caption{~Dynamical spin structure factors from DMRG calculations with different quantum numbers $S_y$ which indicate strength of magnetic fields.
  }
  \label{SI_DMRG}
\end{figure*}

In our notation, $i.e.$, the  Cartesian coordinate system shown in Fig. 1(a) of main text, the direction of the applied magnetic field is parallel to $y$ axis. Since we used finite-size model and the $y$-component of spin quantum number $S_y$ is a good quantum number of our Hamiltonian, the magnetization changes step-wise. Therefore, we compare our experimental data with dynamical structure factor calculated for a given quantum number $S_y$, which can be associated with saturation magnetization as $S_y/S_y^{total} = M/M_{\mathrm{Saturation}}$, with $S_y^{total} = 61/2$ is determined by the size of the cluster for a spin-1/2 system. In other words, we use the quantum number $S_y$ to represent the strength of magnetic field in the DMRG calculations, with $S_y$ = 61/2 corresponds to the saturation field. Thus, on the basis of the fitting magnetization curve in Fig.~\ref{SI_MH}(b), we performed the dynamical DMRG calculations with $S_y$ = 9/2 for $B$ = 2 T, $S_y$ = 23/2 for $B$ = 3 T, $S_y$ = 25/2 for $B$ = 4 T, and $S_y$ = 27/2 corresponds to $B$ = 5 T. Additional calculations with S$_y$ = 17/2, and  S$_y$ = 29/2 were also done to make the evolution of the calculated dynamical structure factors complete. The calculated dynamical structure factors $S(\mathbf{Q}, \omega)$ and its components along the Cartesian axes, $S^{xx}(\mathbf{Q}, \omega)$, $S^{yy}(\mathbf{Q}, \omega)$, and $S^{zz}(\mathbf{Q}, \omega)$ are summarized in Fig.~\ref{SI_DMRG}. As we used isotropic model in our calculations, we have $S^{xx}(\mathbf{Q}, \omega)$ = $S^{yy}(\mathbf{Q}, \omega)$ = $S^{zz}(\mathbf{Q}, \omega)$ for zero field, and $S^{xx}(\mathbf{Q}, \omega)$ = $S^{zz}(\mathbf{Q}, \omega)$ for the non-zero fields. Considering the anisotropy of the $g$ factor, we obtain the total $S(\mathbf{Q}, \omega)$ = $g^2_{ab}$ $S^{xx}(\mathbf{Q}, \omega)$ + $g^2_{ab}$ $S^{yy}(\mathbf{Q}, \omega)$ + $g^2_{c}$ $S^{zz}(\mathbf{Q}, \omega)$. The used the $g$ factors, $g_{ab}$ = 3.25, $g_c$ = 0.3 were measured from ESR experiments [see Fig. 1(b)]. As we have mentioned in the main text, the calculated total $S(\mathbf{Q}, \omega)$ describe our experimental results well. And for the calculations with $S_y$ $\ge$ 17/2 ($i.e.$, in the 1/3 plateau state, see Figs. 3(d)--3(e), and Figs.~\ref{SI_DMRG}), because the intensity of the high-energy diffusive branch in energy range 1 $\textless$ $E$ $\textless$ 1.5 meV (mode IV) mostly comes from the longitudinal (parallel to the applied magnetic field) component $S^{yy}(\mathbf{Q}, \omega)$, and the intensity of the low-energy branches (modes I--III) mostly comes from the transverse components $S^{xx}(\mathbf{Q}, \omega)$, $S^{zz}(\mathbf{Q}, \omega)$, we identify that the mode IV is two-magnon excitation continuum, and modes I--III correspond to the single-magnon excitations~\cite{kamiya2018nature,Chernyshev2009}. It should be noted that the used exchange parameter in the dynamical DMRG calculations is $J$ = 0.48 meV. The $J$ = 0.48 meV is a little smaller than that obtained from the fitting of magnetization data with $J$ = 0.544 meV. This small difference of the exchange coupling parameters between the fitting of magnetization data and the simulation of the INS data is acceptable, since we used such a simple isotropic NN model. We also note that the used quantum numbers $S_y$ were selected on the basis of the fitting curve of magnetization result in Fig.~\ref{SI_MH}(b), and the real strength of the magnetic field corresponds to a selected $S_y$ may have some mismatch with the experimental magnetic field due to the finite-size effects in the magnetization curve for N=61 open cluster. And from the comparison between the experimental data and calculations shown in Fig. 3 of main text, the selected $S_y$ = 23/2 may correspond to a magnetic field above 3 T, and $S_y$ = 25/2 may correspond to a magnetic field above 4 T.
Both the grand canonical DMRG method and the dynamical DMRG technique with an isotropic NN coupling can describe the magnetization data and the INS data well, respectively. This means that the additional interactions including XXZ anisotropy $\Delta$, the next-NN and interlayer couplings are sufficiently smaller than the main NN exchange coupling $J$. We can ignore them safely in this work, although the simulations with some additional interactions may become a little better than what we have now.

\bibliography{csybse2}